\newcommand{\be}{\begin{equation}}
        \newcommand{\ee}{\end{equation}}
        \newcommand{\ba}{\begin{eqnarray}}
        \newcommand{\ea}{\end{eqnarray}}
        \newcommand{\ban}{\begin{eqnarray*}}
        \newcommand{\ean}{\end{eqnarray*}}
         
\newcommand{\R}{{\mathbb R}}

\documentclass{article}
\usepackage {graphicx,latexsym}
\usepackage{amsmath,amsthm}
\usepackage{amsfonts}
\usepackage[utf8]{inputenc}

\usepackage{caption}
\usepackage{subcaption}

\DeclareMathOperator{\Tr}{Tr}
\DeclareMathOperator{\Rp}{Re}
\DeclareMathOperator{\sgn}{sgn}

\begin{document}

\title{
%\letter{
%Classical effective GBFT
Multisymplectic effective\\ 
General Boundary Field Theory
}
\author{
Mona Arjang$^{1, 2}$
and José A. Zapata$^1$
\footnote{e-mail: \ttfamily zapata@matmor.unam.mx} 
\\%%for iop this line is commented%%
%\address{ 
{\it $^1$Centro de Ciencias Matemáticas,}\\ 
{\it Universidad Nacional Autónoma de México} \\ 
{\it C.P. 58089, Morelia, Michoacán, México}\\
{\it $^2$Instituto de Física y Matemáticas,}\\ 
{\it Universidad Michoacana de San Nicolás de Hidalgo,}\\
{\it Edif. C-3, C. U., C.P. 58040, Morelia, Mich., México}
%}
}

\date{}
\maketitle

\begin{abstract} 
The transfer matrix in lattice field theory connects the covariant and the initial data frameworks; 
in spin foam models, it can be written as a composition of elementary cellular amplitudes/propagators. 
We present a framework for discrete spacetime classical field theory in which 
solutions to the field equations over elementary spacetime cells may be amalgamated if they satisfy
simple gluing conditions matching the composition rules of cellular amplitudes in spin foam models.  
Furthermore, the formalism is endowed with 
a multisymplectic structure responsible for local conservation laws. 

Some models within our framework are effective theories modeling a system at a given scale. 
Our framework allows us to study coarse graining and the continuum limit. 
\end{abstract}

\tableofcontents

\section{Classical mechanics on cellular time domains}
% Discrete time classical mechanics à la GBFT
\label{1d}
% - In Hamilton's principle solutions are determined by bdary data 
% $(q_{in}, q_{out}) \in Q \times Q$; thus, it is natural to describe evolution in discrete 
% time steps using $Q \times Q$ to parametrize solutions. 

According to Hamilton's principle 
physical motions of a system evolving in a lapse of time $U\subset \R_t$ are characterized as those histories $q = q(t)$ which are extrema of the action 
\[
S(q) = \int_U {\cal L} (q, \dot{q}) = \int^a_b L(t, q, \dot{q}) dt
\]
when the value of configuration variable is fixed at $\partial U$ to be 
$(q(a), q(b)) \in Q_a \times Q_b$. 
Thus, whenever it is appropriate, it is natural to parametrize solutions by boundary data. 

Additionally, if the interval of interest is 
formed by gluing 
adjacent subintervals 
$U= [a=a_0, a_1] \cup ... \cup [a_{n-1}, a_n=b]$, a solution would be composed of pieces corresponding to each subinterval. A solution in $U$ would have to be a solution in each of the subintervals, and the extra requirements to be an extremum could be interpreted as conditions matching the left and right derivatives at each gluing point. 

The solution in $U$ might be parametrized using $(q(a), q(b)) \in Q_a \times Q_b$, or it could be parametrized by the boundary data of the first subinterval 
$(q(0), q(1)) \in Q_0 \times Q_1$. This can be considered as a sort of initial data characterizing the solution in the whole $U$. Since we are talking about solving the equations of motion without any error, we know that evolution preserves all the structure given by the machinery of symplectic geometry: in particular, evolution 
$e_{ij}: T^\ast Q_i \to T^\ast Q_j$ 
is a symplectic map and 
symmetries have corresponding conserved quantities. If the appropriate map $I_i: Q_i \times Q_{i+1} \to T^\ast Q_i$ were found, initial conditions would be parametrized by $Q_0 \times Q_1$ and the space would have the two form 
$I_0^\ast (dp \wedge dq)$ which in some domain would be non degenerate providing all the elements of a symplectic structure. 
%(In fact, a ``symplectic potential can be imported 
%to the same space using the same map.) 
If we had access to Hamilton's principal function $S_H(q_i, q_{i+1})$, 
defined to be the evaluation of the action in the solution determined by $(q_i, q_{i+1}) \in Q_i \times Q_{i+1}$, 
then the desired map would be given by 
\[
I_i(q_i, q_{i+1}) = \frac{\partial S_H}{\partial q_i}(q_i, q_{i+1}) dq_i . 
\]
After doing the same for the final data (with the aid of the map 
$F_j(q_{j-1}, q_j) = \frac{\partial S_H}{\partial q_j}(q_{j-1}, q_j) dq_j$) we would have devised a way to see evolution in an alternative picture 
$\tilde{e}_{ij}: Q_i \times Q_{i+1} \to Q_{j-1} \times Q_j$ which would 
enjoy of a symplectic structure. 
Here one may complain that we seem to have introduced two different structural two forms $I_i^\ast (dp \wedge dq)$ and 
$F_{i+1}^\ast (dp \wedge dq)$ to each space $Q_i \times Q_{i+1}$, but closer analysis 
(using $ddS_H= 0$) shows us that they agree up to a sign. 

Now, in this picture we can take on the job of investigating the pasting of evolution maps 
$\tilde{e}_{ij}$, $\tilde{e}_{jk}$ to build an evolution for a longer period $\tilde{e}_{ik}$. Since the equations of motion have been solved in each open interval, 
the remaining job is just to paste the final condition of $\tilde{e}_{ij}$ with the initial condition of $\tilde{e}_{jk}$. Proper pasting can be done if we bring this data to 
$T^\ast Q_j$ and match it there. 

This was our preamble to the main part of this section in which we present a version of Veselov's discrete mechanics \cite{Veselov}. The preamble provides a possible interpretation of the framework. We warn the reader that this interpretation looses part of its strength for 
the version of the formalism of Sections \ref{DiscreteScalarGBFT} and \ref{gauge} dealing with 
spacetimes of dimension greater than one because Hamilton's principal function becomes inaccessible when the data at our disposal is the field's value at a discrete set of points. The best that we can do, even in principle, is to postulate a function that approximates Hamilton's principal function from our partial knowledge of the boundary data; for the continuation of this discussion see Section \ref{Reg-Coar-Cont}. 
The preamble also highlights the pasting of subintervals as playing a primary role. Our framework is motivated by the goal of 
having a formalism with 
simple gluing; 
the difference between our version of the framework and Veselov's is that when the system is of the form $L = T-V$ (that is, free theory plus interaction) our pasting conditions are independent of the interaction.

\subsection{Structure emerging from the variational principle}
% The action and its variation

Time has been subdivided into a sequence of cells, lapses of time. 
We picture the time interval of our interest, $U\subset \R_t$, 
as a compact interval divided into a finite collection of 
closed subintervals that we will call {\em time-atoms}. A generic time-atom 
in $U$ is denoted by 
$\nu \in U_{\tiny\mbox{disc}}$, and if it is needed we number such time-atoms $\nu = 1, ..., n$. 

Our discretization is based on decimating each time-atom keeping track only of its 
past boundary $\nu^-\in \nu$, of its future boundary $\nu^+\in \nu$ and of a center point $C\nu\in \nu^\circ$ --representing ``the bulk of $\nu$.'' 
The boundary points of interior time-atoms are shared $\nu^+ = (\nu+1)^-$, and 
$U$'s boundary is composed of $\{1^- , n^+\}$; we write the oriented boundary as 
$\partial U= -1^- + n^+$. 
The system consists of a particle moving in a configuration space $Q$, while our decimated description records its position only in discrete time steps. 
Our record of a history is: 
\[
q= (q(1^-)= q_1^-, ..., q_{n-1}^+=q_n^-, q(Cn)=q_n, q_n^+).
\] 

We are interested in describing histories locally. In the continuum a history is a curve $q: \R_t \to Q$; 
keeping track of time, position and velocity 
$t \overset{\tilde{q}}{\mapsto}(t, q, \dot{q})$ gives us 
the correct arena to study mechanics in the 1st order formalism. 
At the discrete level our 1st order data will be 
the portion of a history at a time-atom 
\[
\nu \overset{\tilde{q}}{\longmapsto} (\nu, q^-, q, q^+) \quad . 
\]
This data gives us the position $q=q(C\nu)$ at the given 
time $C\nu$, and allows us to 
estimate the velocity in two independent ways: 
using $(C\nu, \nu^+) \to (q, q^+)$ or using $(C\nu, \nu^-) \to (q, q^-)$. 
We recall that the map $\tilde{q}$ comes from a history, 
which implies that $\tilde{q}(\nu)$ and $\tilde{q}(\nu +1)$ are not independent: they need to obey $q^+(\nu)= q^-(\nu+1)$. 

Following Veselov \cite{Veselov}, the cornerstone of our version of discrete mechanics is a variational principle and we study the resulting geometrical structures. We start with the action 
\begin{equation}
\label{S}
S(q) = \sum_{\nu \in U_{\tiny\mbox{disc}}} 
L(\tilde{q}(\nu)) \quad ,
\end{equation}
with a discrete lagrangian of the form 
$L(\tilde{q}(\nu))= L^-(\nu, q^-, q) + L^+(\nu, q, q^+)$. 
According to the opening paragraph of this section, the terms in the lagrangian are interpreted as approximations to Hamilton's principal function in the past and future corners of $\nu$ respectively. 

Hamilton's principle determines the motions predicted by our model as the extrema of the action while the end points are kept fixed. 
We will therefore calculate the effect on the action of a variation 
$\delta q = (\delta q_1^-, ..., \delta q_{n-1}^+=\delta q_n^-, \delta q_n, \delta q_n^+)$ of the history $q$. Our notation in the local 1st order format is that 
\[
\delta \tilde{q}(\nu) = \tilde{v}(\nu)=
(v_\nu^- \in T_{q_\nu^-}Q^- , v_\nu \in T_{q_\nu}Q , v_\nu^+ \in T_{q_\nu^+}Q^+)
\] 
is a variation of $\tilde{q}$ at $\tilde{q}(\nu)$. 
Note that the relation $q^+(\nu)= q^-(\nu+1)$ implies 
$dq^+[\tilde{v}(\nu)]= dq^-[\tilde{v}(\nu+1)]$. 
The variation of the action $dS(q) [v]$ 
is a sum of terms over the atoms, but it can be decomposed in a finer manner. This will allow us to separate contributions to the variation into 
bulk terms and boundary terms:
\begin{eqnarray}
\label{dS}
dS(q) [v] &=& 
\frac{\partial L}{\partial q^-} (\tilde{q}(1)) dq^-[\tilde{v}(1)] \nonumber\\
&+&\sum_{\nu \in U_{\tiny\mbox{disc}}}
\frac{\partial L}{\partial q} (\tilde{q}(\nu)) dq[\tilde{v}(\nu)] \nonumber\\
&+&\sum_{\nu, \nu+1 \in U_{\tiny\mbox{disc}}} 
\left(
\frac{\partial L}{\partial q^+} (\tilde{q}(\nu)) 
+ 
\frac{\partial L}{\partial q^-} (\tilde{q}(\nu+1)) 
\right) dq^+[\tilde{v}(\nu)] \nonumber\\
&+& 
\frac{\partial L}{\partial q^+} (\tilde{q}(n)) dq^-[\tilde{v}(n)] \quad .
\end{eqnarray}
From the bulk terms we read two types of equations of motion:
\begin{enumerate}
\item[(i)]
$\frac{\partial L}{\partial q}(\tilde{q}(\nu))=0$, 
equations ensuring that we have a solution in the interior of each time-atom $\nu$ and which do not depend on variables of any other interval.
\item[(ii)]
$\frac{\partial L}{\partial q^+}(\tilde{q}(\nu)) +
\frac{\partial L}{\partial q^-}(\tilde{q}(\nu +1)) = 0$, 
conditions requiring that the final conditions of $\nu$ match with the initial conditions of $\nu+1$ for the time-atoms that are pasted in the interior of $U$. 
\end{enumerate}

We will briefly recall a geometric picture that is standard in mechanics with continuum time before we write its analog in the discrete time framework. 
A history in the continuum $q^{\scalebox{0.7}{\mbox{c}}}$ 
in the first order formalism is seen as the following section 
$U \ni t \overset{\tilde{q}^{\scalebox{0.7}{\mbox{c}}}}{\longmapsto} (t, q(t), \dot{q}(t))$. In this larger space, 
which displays velocities as independent, 
the dynamics of the system is encoded the expression for the variation of the action 
$S^{\scalebox{0.7}{\mbox{c}}}(q^{\scalebox{0.7}{\mbox{c}}}) = \int_{U} {\cal L}^{\scalebox{0.7}{\mbox{c}}}$, written as 
$dS^{\scalebox{0.7}{\mbox{c}}}(q^{\scalebox{0.7}{\mbox{c}}}) 
[\delta q^{\scalebox{0.7}{\mbox{c}}}] = 
-\int_{U} \tilde{q}^{{\scalebox{0.7}{\mbox{c}}}\ast}(\delta \tilde{q} \lrcorner
\hat{\Omega}_{\cal L}^{\scalebox{0.7}{\mbox{c}}}) + 
\int_{\partial U} \tilde{q}^{{\scalebox{0.7}{\mbox{c}}}\ast}(
\delta \tilde{q} \lrcorner \Theta_{\cal L}^{\scalebox{0.7}{\mbox{c}}})$, 
where $\hat{\Omega}_{\cal L}^{\scalebox{0.7}{\mbox{c}}}$ and $\Theta_{\cal L}^{\scalebox{0.7}{\mbox{c}}}$ are defined from this formula and enjoy rich properties; see Chapter 9 of \cite{Arnold}. 
Notice that 
$\delta \tilde{q}$ is the vector field, induced by the variation of the history, that lives 
in the space where the first order formalism takes place. 
We remark that in classical mechanics the variation of the action encodes all the information regarding the dynamics, which in this 
notation is organized as follows: Cartan's form $\Theta_{\cal L}^{\scalebox{0.7}{\mbox{c}}}$ encodes the geometrical structure that arises when the equations of motion, coded in $\hat{\Omega}_{\cal L}^{\scalebox{0.7}{\mbox{c}}}$, vanish. 

A notational warning is in order: the most prominent element of the geometrical structure mentioned above is the symplectic $2$-form arising from evaluating $\Omega_{\cal L}^{\scalebox{0.7}{\mbox{c}}}= -d\Theta_{\cal L}^{\scalebox{0.7}{\mbox{c}}}$ at a component of $\partial U$. Thus, our notation includes $\hat{\Omega}_{\cal L}^{\scalebox{0.7}{\mbox{c}}}$ and 
$\Omega_{\cal L}^{\scalebox{0.7}{\mbox{c}}}$. 
They carry a similar symbol because in a related formulation of mechanics and classical field theory 
they are two parts of the same object. The bundle structure, induces a grading of differential forms considering separately their 
vertical and horizontal parts \cite{VariationalBicomplexnLab}. 
Below we will discretize the horizontal parts and not the vertical parts making $\hat{\Omega}_{\cal L}$ and $\Omega_{\cal L}$ objects that are related by the formalism, but which are different. This digression had the sole purpose of explaining the origin of the notation, and it will not play an essential role in the rest of the article.

Now we develop the discrete analog of this geometrical framework. Our discrete setting suggests keeping the geometrical meaning but instead of integrating differential forms in a domain of spacetime (and its boundary) acting with 
cochains on the domain (and its boundary) seen as a chain composed of elementary cells that we call atoms. 
In this way, the sum over the discrete set $U_{\tiny\mbox{disc}}$ will be written with a notation similar to that of an integral, indicating the action of the cochain on the domain. 
We will write Equation (\ref{dS}) in the form 
\[
dS(q) [v] = - \!\! \sum_{U-\partial U}
\tilde{q}^\ast(\tilde{v} \lrcorner \hat{\Omega}_L) 
+ \sum_{\partial U}
\tilde{q}^\ast(\tilde{v} \lrcorner \Theta_L) \quad . 
\]
Let us explain the notation in detail. 
Starting with the second term, $\partial U$ is the $0$-chain $-1^- + n^+$ in $\R_t$, and 
$\tilde{q}^\ast(\tilde{v} \lrcorner \Theta_L)$ is a $0$-cochain, making 
$\sum_{\partial U}\tilde{q}^\ast(\tilde{v} \lrcorner \Theta_L)= - (\tilde{v} \lrcorner \Theta_L) (q_1^-) + (\tilde{v} \lrcorner \Theta_L) (q_n^+)$. We will define the discrete Cartan form $\Theta_L$ below. Let us explain the first term now: 
since the second term contains the contributions to $dS$ from the boundary degrees of freedom ($q_1^-, q_n^+$), the first term should have all the other terms. This is the meaning of $\sum_{U-\partial U}$; it includes contributions from variations on degrees of freedom $\tilde{q}(\nu)$ for all $\nu\subset U$ except for the boundary degrees of freedom. Generic atoms $\nu$ are seen as $1$-chains, and 
$\tilde{q}^\ast(\tilde{v} \lrcorner \hat{\Omega}_L)$ is a $1$-cochain. 
We will define $\hat{\Omega}_L$ below. 
When $\Theta_L$ and $\Omega_L$ are fed a chain they become $1$-forms acting on variations in the first order format. 
From (\ref{dS}) we can read that 
\[
\Theta_L(\cdot , \tilde{q}(\nu)^-) = 
- \left( \frac{\partial L}{\partial q^-} dq^- \right) (\tilde{q}(\nu))
\quad , \quad 
\Theta_L(\cdot , \tilde{q}(\nu)^+) = 
 \left( \frac{\partial L}{\partial q^+} dq^+ \right) (\tilde{q}(\nu))
\quad , 
\]
\[
\hat{\Omega}_L(\cdot , \tilde{q}(\nu)) = 
-  \left(
\frac{\partial L}{\partial q^-} dq^- +
\frac{\partial L}{\partial q} dq +
\frac{\partial L}{\partial q^+} dq^+ 
\right) (\tilde{q}(\nu))
\quad , 
\]
where the differential $1$-forms displayed in the right hand sides have base point 
$(q_\nu^-, q_\nu, q_\nu^+)\in Q^- \times Q \times Q^+$. 
Here $\tilde{q}^\ast \hat{\Omega}_L$ is seen as having a term corresponding to the bulk of 
$\nu$ and two terms corresponding to $\partial \nu$. 
We caution the reader about a possible confusion generated by our notation: 
The intervals $\nu$ and $\nu +1$ share a point, and we use the convention 
$\tilde{q}(\nu)^+ = \tilde{q}(\nu +1)^-$. The $1$-forms 
$\Theta_L(\cdot , \tilde{q}(\nu)^+)$ and $\Theta_L(\cdot , \tilde{q}(\nu +1)^-)$, however, differ in more than a sign. 

We also define $\Omega_L= - d\Theta_L$; when this object is evaluated on a $0$-chain it becomes a $2$-form acting on variations of a history in the first order format
\[
\Omega_L (\cdot, \cdot, \tilde{q}(\nu)^-) =  \frac{\partial^2 L}{\partial q\partial q^-} dq \wedge dq^-  , \quad 
\Omega_L (\cdot, \cdot, \tilde{q}(\nu)^+) = - 
\frac{\partial^2 L}{\partial q\partial q^+} dq \wedge dq^+  . 
\]
Below we will see that in important circumstances the induced $2$-forms are symplectic. Since $\Omega_L$ induces symplectic forms it is called multisymplectic. 

If the lagrangian in the continuum is of the form of a free theory plus an interaction term 
$L= T-V$, 
the structure of our discretization by decimation with a ``bulk'' representative point and two 
estimates of velocity for each time-atom suggests 
seeking for a discrete lagrangian of the form $L = L^- + L^+$ with 
\begin{eqnarray*}
L^-(\nu, q^-, q) = T(\nu, q^-, q) - \frac{1}{2} V(\nu, q),\\
L^+(\nu, q, q^+) = T(\nu, q, q^+) - \frac{1}{2} V(\nu, q).
\end{eqnarray*}
This feature has two related consequences. The first one is that the equations of motion of type (ii) would be those of the free theory. We say that our model leads to simple gluing of time-atoms. 
The second consequence is that the structural forms 
$\Theta_L$ and $\Omega_L$ defined above are independent of the interaction term.

%%%%

The objects just defined acquire useful properties when only solutions and first variations (variations of the solution that are tangential to the space of solutions inside the space of histories) are considered. In the space of solutions the variation of the action has only boundary contributions; the bulk plays the important role of making the boundary data correlated, but it drops out of the expressions 
\[
dS(q) [v] = \sum_{\partial U}
\tilde{q}^\ast 
(\tilde{v} \lrcorner \Theta_L)=
- \Theta_L(\tilde{v}(1), \tilde{q}(1)^-) + \Theta_L(\tilde{v}(n), \tilde{q}(n)^+) . 
\]

Consider any solution of the equations of motion $q$ and any two 
first variations $v, w$ of it; the identity 
$0= ddS=-\sum_{\partial U}\tilde{q}^\ast \Omega_L$ implies that 
\begin{equation}
\label{Multisymp1d}
\sum_{\partial U} 
\tilde{q}^\ast
(\tilde{w} \lrcorner \tilde{v} \lrcorner
\Omega_L) =0 . 
\end{equation}
This conservation law for the mutisymplectic form is called the multisymplectic formula. In the examples we will see 
that solutions induce a discrete flow; this conservation law implies that the flow preserves the symplectic form induced by 
$\Omega_L$.

Now consider a situation in which a Lie group ${\cal G}$ acts on 
$Q$. 
Then ${\cal G}$ also has an action on histories, 
and an action on the bundle used to describe local first order histories 
$U_{\tiny\mbox{disc}} \times (Q^- \times Q \times Q^+)$ 
by $g(\nu, q^-, q, q^+) = (\nu, g(q^-), g(q), g(q^+))$. 
We consider the case in which 
this group action leaves the discrete lagrangian invariant: 
$L(g(\tilde{q}(\nu))) = L(\tilde{q}(\nu))$ for all $\nu \in U_{\tiny\mbox{disc}}$, for any first order history 
$\tilde{q}$ and for any $g\in {\cal G}$. 
Thus, $S$ is also invariant and the group action preserves the subspace of extrema. This fact, the subspace of extrema containing directions in which $dS$ vanishes leads to conserved quantities. Below we state this discrete analog of Noether's theorem. 

We can use ${\cal G}$'s action on the space of 
1st order histories 
to transform $\xi\in Lie({\cal G})$ into a vector field corresponding to a first variation 
$v_\xi$. Then, for any solution $q$ we have 
\begin{equation}
\label{Noether1d}
0 = dS(q) [v_\xi] = 
\tilde{q}^\ast
\left( \tilde{v}_\xi \lrcorner \Theta_L \right)
|_{\partial U} . 
\end{equation}
Compute 
$J_\xi (\nu^-) = \Theta_L(\tilde{v}_\xi , \tilde{q}(\nu)^-)$ and 
$J_\xi (\nu^-) = \Theta_L(\tilde{v}_\xi , \tilde{q}(\nu)^+)$; 
the equation written above with $U = \nu$ says that in the space of solutions we can write 
$J_\xi (\nu^-)=J_\xi (\nu^+) =J_\xi (\nu)$ obtaining a single quantity that can be associated with the solution at $\nu$. Once we have stablished this, we can look again to Equation (\ref{Noether1d}) as stating that $J_\xi (\nu)$, 
\[
J_\xi (\nu)= 
- \left( \frac{\partial L}{\partial q^-}(\tilde{q}(\nu)) \right) (v_\xi(\nu^-)) , 
\]
is independent of $\nu$; it is a conserved quantity associated with 
the symmetry generator $\xi$. 

In the examples we will see that the conservation law for $\Omega_L$ implies that there is an induced symplectic form that is conserved, and we will apply our discrete version of Noether's theorem.

\subsection{Comments and 
discrete time examples}
\label{1dExamples}
%\vskip0.3cm
%\noindent
%{\em Further structures}\\
Here we mention structures that will be introduced in the examples given in this subsection, and that will be 
treated at length when we study field theories on discrete spacetimes. 

The example of {\em a particle on a potential} may be seen as a preamble to our treatment of spacetime scalar degrees of freedom, and the example of {\em rigid body motion} may serve as a preamble to our treatment of gauge fields. 

The reader may choose to go directly to the field theory sections. 

\vskip0.3cm
\noindent
{\em Reduction by solving the gluing equations}\\ 
We may be interested in solving the gluing equations to eliminate boundary variables. 
This reduction can be done explicitly because the gluing equations are simple. 
The resulting reduced formalism is described by an action principle with reduced lagrangian defined as the the value of the non-reduced lagrangian on 
histories that solve the gluing equations. Note that, in several cases of interest,
the gluing equations admit multiple solutions. 
See Section \ref{Reduced} and the examples of this section 
for a detailed discussion. 

\vskip0.3cm
\noindent
{\em A formalism in terms of atomic boundary data}\\ 
Another attractive objective is to solve the interior equations of motion to 
eliminate the ``atomic bulk variables'' and have a formalism where the only 
data sits on the boundaries between neighboring time-atoms (subintervals), and the only 
equations of motion are gluing equations. In general, this reduction leads to a complicated reduced lagrangian. Thus, the resulting gluing equations are complicated as are the expressions for the structural forms giving conservation laws. The obvious alternative is to propose an effective lagrangian in terms of discrete histories resulting from a decimation only at the boundaries of the time-atoms. The resulting formalism 
falls within the framework of Veselov's discrete time mechanics \cite{Veselov}. 
The field theory analog of this boundary data formalism is discussed in 
Section \ref{BdaryDataF}. 

\vskip0.3cm
\noindent
{\em Covariant hamiltonian picture}\\ 
A hamiltonian picture would have the advantage of being defined on a collection of cotangent bundles with canonical symplectic structures. Veselov's discrete mechanics starts with a lagrangian variational principle and develops an equivalent hamiltonian picture \cite{Veselov}. We introduce the basic notions of a covariant hamiltonian picture for field theories on discrete spacetimes in 
Section \ref{Canonical}.

\vskip0.3cm
\noindent
{\em Example 1. Particle moving on euclidian space with a potential}\\ 
In our first example the discrete lagrangian is $L=L^- + L^+$ with 
$L^-(\tilde{q}(\nu))= [\frac{m}{2}(\frac{q_\nu-q_\nu^-}{a})^2 - V(q_\nu)]a$ and 
$L^+(\tilde{q}(\nu))= [\frac{m}{2}(\frac{q_\nu^+-q_\nu}{a})^2 - V(q_\nu)]a$, 
where the points $v^-, C\nu , \nu^+$ are equally spaced and the lapse between them has been denoted by ``$a$.'' 

For every interval $\nu \subset U$ we have an 
equation of motion associated with variations of the history at its ``bulk point'' 
\[
0 = dS(q) [\frac{\partial }{\partial q^A_\nu}] = 
2a[-g_{AB} \frac{m}{2a} \left( \frac{q_\nu^+-q_\nu}{a} - \frac{q_\nu-q_\nu^-}{a} \right)^B
- \frac{\partial V}{\partial q^A}(q_\nu)] .
\]
There are also equations of motion related to variations of the history over points 
$\nu^+=(\nu+1)^-$ 
where two intervals $\nu, (\nu+1)\subset U$ meet 
\[
0 = dS(q) [\frac{\partial }{\partial q^{+A}_\nu}] = 
g_{AB} \frac{m}{a} \left( (q_\nu^+-q_\nu)- (q_{\nu+1}-q_{\nu+1}^-) \right)^B .
\] 
These equations are independent of the potential --{\em they would be the same for a free particle}-- and they simply imply that $q_\nu^+-q_\nu = q_{\nu+1}-q_{\nu+1}^-$. We call them simple gluing equations. 

Consider any solution $q$ of the equations of motion, as we saw 
$dS(q) [v] = -\Theta_L(\tilde{v}(1), \tilde{q}(1)^-) +
\Theta_L(\tilde{v}(n), \tilde{q}(n)^+)$ with 
\[
\Theta_L(\cdot , \tilde{q}(\nu)^-) =  m g_{AB} \frac{(q_\nu-q^-_\nu)^B}{a} dq^{-A} 
\quad ,
\]
\[
\Theta_L(\cdot , \tilde{q}(\nu)^+) =  m g_{AB} \frac{(q^+_\nu-q_\nu)^B}{a} dq^{+A}
\quad . 
\]
Now we will see a concrete example of the conservation law (\ref{Multisymp1d}) for 
$\Omega_L = -d\Theta_L$. 
For any pair of first variations $v, w$ of the solution $q$ 
\begin{eqnarray*}
0 &=& v\lrcorner w \lrcorner ddS(q) = \Omega_L(\tilde{v}(1), \tilde{w}(1), \tilde{q}(1)^-) 
- \Omega_L(\tilde{v}(n), \tilde{w}(n), \tilde{q}(n)^+) \\
&=& - \frac{m}{a} g_{AB} dq^B \wedge dq^{-A} (\tilde{v}(1), \tilde{w}(1))
- \frac{m}{a} g_{AB} dq^B \wedge dq^{+A}(\tilde{v}(n), \tilde{w}(n))
\end{eqnarray*}
An equation of this type also holds for any ``subsolution'' 
$q|_{[\nu^-, \nu^+]}$, which means that for every solution $q$ and 
at each time time-atom $\nu$ 
we have a two form 
$\Omega_L(\tilde{v}(\nu), \tilde{w}(\nu), \tilde{q}(\nu)^-) 
= \Omega_L(\tilde{v}(\nu), \tilde{w}(\nu), \tilde{q}(\nu)^+)$. Moreover, this two form is closed and non degenerate, which makes it a symplectic form. 
If at each time-atom we parametrize solutions by the initial data, 
(\ref{Multisymp1d}) may be written as the conservation of the symplectic form 
\[
- \frac{m}{a} g_{AB} dq^B \wedge dq^{-A} (\tilde{v}(1), \tilde{w}(1))=
- \frac{m}{a} g_{AB} dq^B \wedge dq^{-A} (\tilde{v}(n), \tilde{w}(n)) \quad .
\] 

Notice that this structural equation does not involve the potential $V$. The potential determines the set of solutions and the relation between initial data and solutions, but the form of the equation expressing the conservation of the symplectic form is independent of the potential --{\em it would be the same for a free particle}--. 

Now consider the case of a free particle, $V=0$. In this case the action is invariant under uniform translations. A first variation of this type in the first order format is of the form $\tilde{v}(\nu) = (v^A \frac{\partial }{\partial q^{-A}_\nu}, 
v^A \frac{\partial }{\partial q^A_\nu}, v^A \frac{\partial }{\partial q^{+A}_\nu})$ for every $\nu \subset U$. According to our version of Noether's theorem (\ref{Noether1d}), for any solution $q$ and any $v \in \R^3$ we have 
$0= dS(q) [v]= -\Theta_L(\tilde{v}(1), \tilde{q}(1)^-) +
\Theta_L(\tilde{v}(n), \tilde{q}(n)^+)$. 
This is the conservation of linear momentum in this discrete time setting, and 
it implies that 
\[
q_1-q^-_1 = q^+_n-q_n \quad .
\]

Before moving to another example, we comment on the relation between our version of discrete time mechanics and other versions of Veselov's discrete time mechanics (see for example \cite{MarsdenEtAl}). Our discretization of the time axis has two types of points: those labeled by $C\nu$ for some $\nu$ and those in the middle of $C\nu$ and $C(\nu+1)$. 
All other versions of discrete time mechanics use a simple discretization of time with all the events being of the same type. 
Below we will ``erase'' half of the points in our discretization by solving the equations of motion associated with them. This is the $1$-dimensional case of the reduced formalism described in Section \ref{Reduced}. 

Recall that in our discrete lagrangian, the potential $V$ is independent of 
$q_\nu^+=q_{\nu+1}^-$ for $\nu, \nu +1 \subset U$. Because of this, the equations of motion associated with those degrees of freedom are very simple: the gluing equations 
$q_\nu^+-q_\nu = q_{\nu+1}-q_{\nu+1}^-$. 
We will solve this equations to obtain a reduced model with fewer variables. 
The boundary variables will be eliminated and the remaining variables will be the $q_\nu$s, the ones that 
represent the position at 
the ``bulk'' of the time-atoms. In order to write the reduced model properly we consider regions of a new type; regions $U'$ whose boundary is located where the remaining degrees of freedom sit, $\partial U' = \{ Ck, C(k+N) \}$ with $1 \leq k < k+N \leq n$. 
Consider a history $q$ that is a solution of the gluing equations of motion in a region $U' \subset U$ of the described type. 
Now let us evaluate the action on a history $q$ which solves the gluing equations of motion restricted to $U'$. 
This portion of the history is parametrized by the remaining variables and 
will be written as 
$q^r= (q_k, \ldots , q_{k+N})$. 

We obtain a reduced model with action 
\[
S^r(q^r)= S(q|_{U'}) =  \!
\sum_{i=0}^N 
\left( \frac{m}{2}(\frac{q_{k+i+1}-q_{k+i}}{2a})^2 - 
\frac{V(q_{k+i}) +V(q_{k+i+1})}{2} \right) \! 2a . 
\]
This action is much closer to the one used in other approaches. 
The equations of motion resulting from $S_r$ 
are exactly those equations of motion of $S$ that we have not solved. 
Thus, an extremum $q$ of $S$ yields an extremum of $S_r$. Conversely, if $q^r$ is an extremum of $S_r$ we can generate from it an extremum $q$ of $S$ restricted to a region $U'' \subset U'$ of the allowed type. We simply set $q_\nu=q^r_\nu$ for every $\nu \subset U''$, and solve the gluing equations by defining 
$q_\nu^+= q_{\nu+1}^- = \frac{1}{2} (q^r_\nu + q^r_{\nu+1})$ if $\nu, \nu+1 \subset U''$. 
This reduced model also comes with a geometric structure that can be derived from the variation of the action. A key difference is that the formula for the conservation of the symplectic structure involves the potential explicitly. 

It would also be interesting to study 
a possible formalism that used only boundary variables $q_\nu^-, q_\nu^+$ for each atom $\nu$. In general cases, building this formalism as a reduction of our original model leads to complicated lagrangians, but we can also define a new regularization with this, simpler, time discretization. A reasonable regularization is 
$S^b(q^b) = \sum_U
\left(  \frac{m}{2}(\frac{q_\nu^+-q_\nu^-}{2a})^2 - 
V(\frac{q_\nu^- + q_\nu^+}{2}) \right) 2a$.

The comparison between the formalisms derived from $S$, $S^r$ and $S^b$ may be summarized in the following list: 
\begin{enumerate}
\item
The models defined by $S$ and $S^r$ are equivalent in the sense that their solutions are in one-to-one correspondence. 
\item
The models of $S$ and $S^r$ are of similar complexity in the sense that the operation 
needed to find solutions of one from solutions to the other is trivial. 
\item
The model defined by $S^b$ is not equivalent to the other formalism, but in the continuum limit they should all agree. 
\item
The model of $S^b$ has the interpretation of being an approximation to the model obtained by reducing by solving the bulk variables; only the solution in the interior of each $\nu$ has been replaced by free motion. The reduced model by solving the interior equations of motion would be equivalent to the models of $S$ and $S^r$. 
\item
The model of $S^b$seems to be of similar complexity to that of $S^r$. 
\item
The geometric structure 
in the space of histories 
leading to a conserved symplectic structure and Noether's theorem is essentially simpler in the model of $S$, since it is independent of any potential. 
\end{enumerate}

\vskip0.3cm
\noindent
{\em Example 2. Discrete time models for rigid body motion} \\
Our second example models a rigid body in $N$-dimensional euclidian space moving freely and described from the center of mass frame. 
This example will serve as a preamble to our treatment of gauge fields in Subsections \ref{gauge} and \ref{BF+subsection}.

The variables used to describe histories include $SO(N)$ group elements giving decimated information about the angular position of the body. 
Euclidean reference frames are fixed in the body and in the laboratory: the are called the ``body frame'' and the ``space frame.'' After an auxiliary euclidian isometry between the two frames is set, the angular position of the body with respect to space is prescribed by moving the auxiliary isometry with a $SO(N)$ group element. 

The variable $q_\nu$ gives a map from the body frame to the space frame telling us the angular position of the body with respect to space at time $C\nu$; similarly, $q_\nu^+$ gives information about the angular position of the body with respect to space at time $\nu^+$, and 
$q_\nu^-$ stores similar information corresponding to time $\nu^-$. 

The angular displacement from time $C\nu$ to time $\nu^+$ according to the body frame is 
$\Phi_\nu^+ = q_\nu^{+ \, T} q_\nu \in SO(N)$; this object lets us estimate the angular velocity in the body frame in the time-atom $[ C\nu, \nu^+ ]$. 
Notice that 
if the auxiliary space frame had been chosen differently and had another orientation, the same history would be described by left-translated variables $g q_\nu, g q_\nu^+$, etc (for some $g \in SO(N)$); however, the angular displacement in the body frame would not be affected by such time independent left-translations. 

We will also use variables in the Lie algebra that are related to angular momentum; the exact relation to angular momentum will be given below after we derive it using our formalism. 
As we did before we will divide time atoms in two segments; in this example we will use the notation $\nu = s_\nu^- \cup s_\nu^+$, and 
the Lie algebra variables will be denoted by $e_{s_\nu^-} , e_{s_\nu^+} \in so(N)$. We remark that these variables are in the body frame; unfortunately, we use lower case letters (as opposed to the convention of using capital letters for body frame objects \cite{Arnold}) 
because the notation is compromised by the goal of writing this model 
in a way that shows its similarities with the models of 
modified BF theories presented in Subsection \ref{BF+subsection}.
Notice that since our $so(N)$ ``$e$'' variables are associated to the body frame, they 
do not change under rotations of the fiducial space frame. 

In the first order format a history is described by 
\[
\tilde{qe}(\nu)= (q_\nu^-, q_\nu , q_\nu^+, e_{s_\nu^-} , e_{s_\nu^+}), 
\]
where the first three entries belong to $SO(N)$ and the last two to $so(N)$. 
We will give more explanations about the variables used above after we write the lagrangian for the model. 
Consider a history displayed in first order format $\tilde{qe}$; 
it may be modified by a left-translation resulting in the new 
first order history 
$\nu \longmapsto (lq_\nu^-, lq_\nu, lq_\nu^+, e_{s_\nu^-} , e_{s_\nu^+})$ 
for all $\nu \subset U$. 
Let $v_\xi^L$ be a generator of variations of this type. The calculations are simplified if we write 
$v_\xi^L= \frac{\rm d}{\rm ds}|_{s=0}q$ with $q(s)= l(s) q$ where $l(0) = \mbox{id}$ and 
$\dot{l}(0)=\xi \in so(N)$. 
In the 1st order format we can write 
$\tilde{v}_\xi^L(\nu)=(\xi q_\nu^-, \xi q_\nu, \xi q_\nu^+, 0, 0)$. 
For general variations we will write 
$\delta\tilde{qe}(\nu)= \tilde{v}_\xi^L(\nu)= 
(\xi_\nu^- q_\nu^-, \xi_\nu q_\nu ,  \xi_\nu^+ q_\nu^+, \xi_\nu^{m-}, \xi_\nu^{m+})$ or 
$\tilde{v}_\xi^R(\nu)= 
(q_\nu^- \xi_\nu^-, q_\nu \xi_\nu , q_\nu^+ \xi_\nu^+, \xi_\nu^{m-}, \xi_\nu^{m+})$. 

In our formalism the geometrical structure is derived after a separation of these local degrees of freedom into the set of those interior to $\nu$ and those associated with $\partial \nu = -\nu^- + \nu^+$. In this case $q_\nu$ , $e_{s_\nu^-}$ and $e_{s_\nu^+}$ are considered interior to $\nu$, while $q_\nu^-= q_{\nu -1}^+$ is associated with $\nu^-$, and 
$q_\nu^+=q_{\nu +1}^-$ is associated with $\nu^+$. 
The action may be written as 
$S(qe)= \sum_{\nu \in U_{\tiny\mbox{disc}}} L(\tilde{qe}(\nu))$ with discrete lagrangian of the form $L=L^- + L^+$ with 
\[
L^-(\tilde{qe}(\nu))= 
e_{s_\nu^- \, i} \theta_\nu^{- \, i} 
- \frac{a}{2} I^{-1 \, ij} e_{s_\nu^- \, i} e_{s_\nu^- \, j} , 
\]
\[
L^+(\tilde{qe}(\nu))= 
e_{s_\nu^+ \, i} \theta_\nu^{+ \, i} 
- \frac{a}{2} I^{-1 \, ij} e_{s_\nu^+ \, i} e_{s_\nu^+ \, j} , 
\]
where 
$\theta_\nu^{- \, i} = - \Tr( f^i \Phi_\nu^-)$ and 
$\theta_\nu^{+ \, i} = - \Tr( f^i \Phi_\nu^+)$ with 
$\Phi_\nu^- = q_\nu^{- \, T} q_\nu$ and 
$\Phi_\nu^+ = q_\nu^T q_\nu^+$. 
We have used $I$ to denote the moment of inertia tensor in the body frame 
that can be assumed to be a diagonal invertible matrix. 
We consider $so(N)$ as the subspace of skew-symmetric $N$ by $N$ matrices 
with inner product $(A, B) = \Tr(A^T B)$. We have chosen 
an orthonormal basis $\{ f_i \}$. 
Again we remark that 
$\theta_\nu^-, \theta_\nu^+$ are body frame objects even when we are using lower case letters. In this notation we will frequently find expressions of the form $-\Tr(f^i A)$, which is ``the $i$th component'' of 
$\frac{1}{2}(A - A^T)$, the skew-symmetric part of $A$. Thus, if the lapse $[C\nu, \nu^+]$ is short with respect to the speed of rotation, $\theta_\nu^+$ approximates the angular velocity. 

A brief remark about our first order history and our notation may help now. It may seem odd that $\tilde{qe}(\nu)$ has three $q$ variables and only two $e$ variables. It would be more appropriate to see two pairs of $q$ variables which enable us to define the angular displacement $\Phi_{s^-}$ during the past segment and the angular displacement 
$\Phi_{s^+}$ during the future segment. The lagrangian does not contain discrete velocities (or displacements) of the $e$ variables. The first term of this 1st order 
lagrangian makes the $e$ variables seeds for the momenta of angular displacements, and that is why we have a single $e$ variable (for each of the segments of a time-atom). 

The most important feature of this model is the invariance of its action under rotations of the fiducial space frame: invariance of the action under time independent left-translations of our $SO(N)$ ``$q$'' variables. 
We know that symmetries give us conserved quantities using the formula 
$0=dS(qe)[v^L_\xi] = -\Theta(\tilde{v}^L_\xi , \tilde{qe}(1^-)) + 
\Theta(\tilde{v}^L_\xi , \tilde{qe}(n^+))$, that holds when $qe$ is a solution and $v^L_\xi$ is the generator of the symmetry. In this case the symmetry 
is time independent left-translations, and the only boundary variables are $q_1^-$, $q_n^+$. For the past boundary we calculate 
\[
\Theta(\tilde{v}^L_\xi , \tilde{qe}(1^-))= - 
e_{s_1^- \, i}  \frac{\partial}{\partial q^-}
\theta^{- \, i}(1^-) dq^-[\xi q^-_1 ] 
= - \Tr(e_{s_1^-} q_1^{- \, T} \xi q_1) = \xi^j m_{1 \, j}^-, 
\]
where 
$m_{1 \, j}^-= -\Tr(f_j q_1e_1^- q_1^{- \, T})$ is the space angular momentum at $1^-$; and for the future boundary the analogous calculation yields 
$\Theta(\tilde{v}^L_\xi , \tilde{qe}(n^+))= \xi^j m_{n \, j}^+$, where 
$m_{n \, j}^+= -\Tr(f_j q_n^+e_n^+ q_n^T)$. 
Since the conservation law holds for any subregion of $U$, we 
obtain a $so(N)$ element that may be evaluated at any $\nu$ and represents the space angular momentum of the solution $qe$: 
\[
\frac{1}{2} (q_\nu e_{s_\nu^-} q_\nu^{- \, T} +  q_\nu^- e_{s_\nu^-} q_\nu^T)
=m_\nu^- (qe)= m_\nu^+ (qe)= 
\frac{1}{2} (q_\nu^+ e_{s_\nu^+} q_\nu^T + q_\nu e_{s_\nu^+} q_\nu^{+ \, T}) .
\]
The body angular momentum will have several names depending on where it is evaluated. We define 
\[
u_\nu^- = q_\nu^{- \, T}  m_\nu^- q_\nu^- , \quad 
u_\nu^+ = q_\nu^{+ \, T}  m_\nu^+ q_\nu^+ , \quad 
w_\nu^-= q_\nu^T m_\nu^- q_\nu ,  \quad 
w_\nu^+= q_\nu^T m_\nu^+ q_\nu . 
\]
The body angular momentum will appear in the variation of the action written in another form.%
\footnote{
$u_{\nu \, j}^- \!\! = \! -\Tr(f_j \Phi_\nu^- e_{s_\nu^-}), \, \, 
u_{\nu \, j}^+ \!\! = \!  -\Tr(f_j e_{s_\nu^+} \Phi_\nu^+), \, \, 
w_{\nu \, j}^- \!\! = \!  - \Tr(f_j e_{s_\nu^-} \Phi_\nu^-), \, \, 
w_{\nu \, j}^+ \!\! = \!  - \Tr(f_j \Phi_\nu^+ e_{s_\nu^+})$.
} 

Now we write the variation of the action for arbitrary variations of the history.  
Since the action is written in terms of objects in the body frame, its variation is easier to calculate using right-invariant vector fields. 
We obtain 
\[
dS(qe) [v^R_\xi] = 
- \sum_{U-\partial U}\tilde{qe}^\ast(\tilde{v}^R_\xi \lrcorner \hat{\Omega}_L) 
+ \sum_{\partial U} \tilde{qe}^\ast(\tilde{v}^R_\xi \lrcorner \Theta_L) , 
\]
with 
\[
\Theta_L(\tilde{v}^R_\xi, \tilde{qe}(\nu^-)) = 
 -\Tr(e_{s_\nu^-} \xi_\nu^- \Phi_\nu^- ) =
\xi_\nu^{- \, j} u_{\nu \, j}^- , \quad 
\Theta_L(\tilde{v}^R_\xi, \tilde{qe}(\nu^+)) = \xi_\nu^{+ \, j} u_{\nu \, j}^+ , 
\]
\begin{eqnarray*}
\hat{\Omega}_L(\tilde{v}^R_\xi, \tilde{qe}(\nu)) &=&
\xi_\nu^{- \, j} u_{\nu \, j}^- - \xi_\nu^{+ \, j} u_{\nu \, j}^+ 
- \xi_\nu^j (w_{\nu \, j}^- - w_{\nu \, j}^+) \\
&&- \xi_{\nu \, i}^{m\, -} (\theta_\nu^{- \, i} - a I^{-1 \, ij} e_{\nu \, j}^-) 
- \xi_{\nu \, i}^{m\, +} (\theta_\nu^{+ \, i} - a I^{-1 \, ij} e_{\nu \, j}^+) . 
\end{eqnarray*}

The first equations of motion which follow from our expression for $\hat{\Omega}_L$ 
are conditions on the matching of body angular momentum: 
\begin{itemize}
\item
the condition at $C\nu$ is $w_\nu^- = w_\nu^+$; 
\item
the gluing condition at $\nu^+$ is 
$u_\nu^+ = u_{\nu +1}^-$. 
\end{itemize}
Together with the identities 
\[
w_\nu^- = \Phi_\nu^{- \, T}  u_\nu^- \Phi_\nu^- \quad , \quad 
u_\nu^+ = \Phi_\nu^{+ \, T}  w_\nu^+ \Phi_\nu^+ , 
\]
these equations of motion are equivalent to the condition expressing that the space angular momentum is conserved. 

The remaining equations of motion link body angular momentum with body angular velocity 
\[
e_{s_\nu^- \, i} =  I_{ij}  \frac{\theta_\nu^{- \, j}}{a} \quad , \quad 
e_{s_\nu^+ \, i} =  I_{ij} \frac{\theta_\nu^{+ \, j}}{a} . 
\]
For example, these equations imply 
$w_{\nu \, j}^-= - \Tr(f_j f^i \Phi_\nu^-)I_{ik} \frac{\theta_\nu^{- \, k}}{a}$. 
Notice that if the body angular momentum $w_\nu^-$ is known, 
this equation does not uniquely determine the angular displacement $\Phi_\nu^-$ 
because the relation between $\Phi_\nu^-$ and $\theta_\nu^-$ is, for generic points, two to one.

In this model the multisymplectic form $\Omega_L=- d \Theta_L$ is 
\begin{eqnarray*}
\Omega_L(\tilde{v}_\xi^R, \tilde{w}_\eta^R, \tilde{qe}(\nu^-)) =
\!\!\!\! &-& \!\!\!\!\tilde{v}_\xi^R[\Theta_L(\tilde{w}_\eta^R, \tilde{qe}(\nu^-))] 
+\tilde{w}_\eta^R[\Theta_L(\tilde{v}_\xi^R, \tilde{qe}(\nu^-))] \\
&+& \!\!\!\! \Theta_L([\tilde{v}_\xi^R, \tilde{w}_\eta^R], \tilde{qe}(\nu^-)) \\
=
\!\!\!\! &-&\!\!\!\! ( \eta_\nu^{-\, j} \tilde{v}_\xi^R(u_{\nu \, j}^-) 
- \xi_\nu^{-\, j} \tilde{w}_\eta^R(u_{\nu \, j}^-)) (\tilde{qe}(\nu)) ,  
\end{eqnarray*}
\[
\!\!\!\!\!\!\!\!\!\!\!\!\!\!\!\!\!
\Omega_L(\tilde{v}_\xi^R, \tilde{w}_\eta^R, \tilde{qe}(\nu^+)) =
- ( \eta_\nu^{+\, j} \tilde{v}_\xi^R(u_{\nu \, j}^+) 
- \xi_\nu^{+\, j} \tilde{w}_\eta^R(u_{\nu \, j}^+)) (\tilde{qe}(\nu)). 
\]
Thus, for any first variations $v, w$ of any solution $qe$ the multisymplectic formula,  
$0= - \Omega_L(\tilde{v}, \tilde{w}, \tilde{qe}_1^-) 
+ \Omega_L(\tilde{v}, \tilde{w}, \tilde{qe}_n^+)$ is explicitly written as 
\[
0= 
( \eta_1^{-\, j} \tilde{v}_\xi^R(u_{1 \, j}^-) 
- \xi_1^{-\, j} \tilde{w}_\eta^R(u_{1 \, j}^-)) 
- 
( \eta_n^{+\, j} \tilde{v}_\xi^R(u_{n \, j}^+) 
- \xi_n^{+\, j} \tilde{w}_\eta^R(u_{n \, j}^+)).
\] 
Since this conservation law is valid for any subregion, we can write it 
in terms of the initial data of each interval, as we did above for the free particle. We obtain the conservation of the symplectic structure  
\[
\eta_1^{-\, j} \tilde{v}_\xi^R(u_{1 \, j}^-) 
- \xi_1^{-\, j} \tilde{w}_\eta^R(u_{1 \, j}^-) 
=
\eta_n^{-\, j} \tilde{v}_\xi^R(u_{n \, j}^-) 
- \xi_n^{-\, j} \tilde{w}_\eta^R(u_{n \, j}^-)  \quad .
\]

%%%%%%%%%%%
Now we will comment on reduced versions of this model. 
The only equations that are simple to solve are $e_{s^-}= I \frac{\theta^-}{a}$, 
$e_{s^+}= I \frac{\theta^+}{a}$, and the $e$ variables can be eliminated. 
The second thing to notice is that we are modeling free motion; the differences with the particle in euclidian space have origin in the configuration space not being flat (with respect to the metric given by the kinetic energy) 
and the invariances of the action coming from a non abelian group. 
Thus, reduction by solving the gluing equations and reduction by solving the interior equation of motion are equivalent. 
A qualitative difference with our previous example that is due to the discreteness of the time parameter, 
is that the equations of motion have multiple solutions. Consider the simple case when the inertia tensor $I$ is a multiple of the identity; in the continuum time description the object moves with uniform angular velocity according to the body frame (and also the space frame). Thus, angular displacement during motion for a given lapse of time is independent of the starting time. 
Discrete motion with equal angular displacement in the time-atoms $[\nu^-, C\nu]$ and 
$[C\nu , \nu^+]$ 
is a solution to the equations of our model, but we saw that generically there is another solution. 
The gluing equations deal only with the matching of body angular momentum as measured as outgoing at the future boundary of a time-atom and measured as incoming at the past boundary of the neighboring time-atom. This equation, as a condition on momentum, 
has unique solutions; however, when the equation is written as a condition on 
angular displacement, double solutions arise.

%%%%%%

%%
Veselov \cite{Veselov} proposed a model describing rigid body motion 
whose discrete lagrangian may be rewritten as 
$L=L^- + L^+$ with 
\[
L^-(\tilde{q}(\nu))= \frac{2N}{a}[1 - \frac{1}{N}\Tr(q_\nu^-Iq_\nu^T)] , 
\]
and the analogous $L^+$; where $I$ is a diagonal matrix, and $M^T$ denotes 
the transpose of the matrix $M$. 
We have normalized and reversed the sign in the trace in a way that physical motions correspond to minima of the action and also with the purpose of simplifying the study of the continuum limit. An extra bonus of this disguise for Veselov's action is that the analogy with Wilson's lattice gauge theory action becomes even more apparent. 
The reader is invited to study this action and re-derive Veselov's discrete time model for rigid body motion. The invariance of the action under left-translations induces a conservation law containing all the information present in the equations of motion which in a continuum limit recover the Euler-Arnold equations. 
It would be a good preamble to our 
lattice gauge theory example presented in Subsection \ref{lgt}. 

Our model and Veselov's model are not equivalent. In both of them space angular momentum is conserved, but the relation between angular momentum and angular velocity agrees only up to first order in angular velocity.

% 3) Discrete spacetime classical GBFT
% Lagrangian GBFT on discrete spacetimes
\section{Local lagrangian field theory 
on discrete\\ 
spacetime}
%\section{Discrete spacetime classical GBFT}
\label{framework}
Let us consider a field theory where histories are local sections 
$M\supset U \overset{\phi}{\longrightarrow} E$
of a bundle over spacetime, $E \overset{\pi}{\longrightarrow} M$ with standard fiber 
${\cal F}\approx \pi^{-1} (x)$. 
Physical motions are selected by Hamilton's principle: possible motions are those histories which are extrema of the action
\[
S(\phi) = \int_U {\cal L} (j^1\phi) 
\]
when the field is kept fixed at $\partial U$. In this first order framework for classical field theory the lagrangian density is a function of the history and its partial derivatives. The symbol $j^1\phi$ denotes an equivalence class of sections up to first order. Two sections at point $x\in M$ are equivalent if they agree at zeroth order ($\phi_1(x) = \phi_2(x)$), and they also agree up to first order as we depart from $x$ (which means that their partial derivatives agree). 
In a local trivialization we have 
$x \overset{j^1\phi}{\longmapsto}(x, \phi, \partial{\phi})$. 
This is the essence of the first jet bundle. 
Starting from this variational principle and using this language, the framework of multisymplectic field theory 
may be developed as done in \cite{MarsdenEtAl}. 

A similar study may be carried out for gauge fields.
In this class of field theories histories are connections on a principal $G$-bundle over a region of spacetime $P  \overset{\pi}{\longrightarrow} U \subset M$; and the lagrangian densities that we consider depend only on connections modulo bundle equivalence maps, allowing us to consider the physical degrees of freedom to be connections modulo internal gauge transformations. However, 
working at the level of connections, while looking for physically meaningful expressions among the gauge invariant functions, is usually more convenient. 
The starting point is the variational problem with action 
\[
S(A) = \int_U {\cal L} (j^1A) . 
\]
Only gauge invariant actions are physically acceptable. 
For a multisymplectic treatment of this type of theories see \cite{GIM}. 

We will also study models  
which include a Lie algebra valued $2$-form 
and possibly a spacetime scalar 
as degrees of freedom. We will consider actions of the type 
\[
S(A, B, \varphi) = \int_U {\cal L} (j^1A, B, \varphi) =
\int_U (B\wedge F + \Phi(B, \varphi)). 
\]
Actions in this family have very interesting properties, but the main reason behind the interest in them is that the family includes Plebanski's action for general relativity \cite{BF, Plebanski, KModified}. 

In this section we will develop a framework for multisymplectic field theory on a discrete spacetime; the principal ideas behind this framework were introduced by Marsden, Patrick and Shkoller in their work on variational integrators \cite{MarsdenEtAl}, but our variation 
of the framework uses a more structured discretization which will be advantageous in some contexts. 
Spacetime is presented as a collection of 
structured pieces or atoms. The structure in each atom 
allows for a separation of the degrees of freedom into 
``bulk'' and ``boundary'' degrees of freedom as shown in our one-dimensional warm-up in Section \ref{1d}. 
Also in parallel to the $1$-dimensional case, 
we will see that the 
structured spacetime atoms 
provide a natural notion of discrete sections that ``agree up to first order'' inside them. We hope that this discrete analog of the first jet bundle that allows for a first order local formalism will be useful in a wide range of  applications. Another important property of this discretization is that it leads to 
a division of the field equations into some that are internal to the atoms and other equations that are read as ``simple gluing'' conditions for local solutions along the shared boundaries. 
In the case of gauge theories, the discretization of spacetime and the variables were introduced by Reisenberger 
\cite{Reisenberger}, and they are core ingredients behind spin foam models \cite{SF}. 
However, the multisymplectic structure that we add to this framework 
has not been published before. 
Our study includes pure gauge fields, and it produces a lattice gauge theory. 
We remark that this lattice gauge theory differs from most because we do not use a 
regular lattice; however, if we solve our simple gluing conditions, we arrive at a reduced model (see Subsection \ref{lgt}) defined on an ordinary lattice whose action is closely related to Wilson's action. 
The structured discretization was used previously in quantum lattice gauge theory (see for example \cite{Reisenberger94, ArocaGambiniFort}), but its ability to cleanly separate bulk and boundary degrees of freedom leading to a multisymplectic structure for classical lattice gauge theory 
had not been exploited before. 
Our study provides a relation between lagrangian and hamiltonian frameworks of classical lattice gauge theories on discrete spacetimes that in a continuum time limit reproduces the formalism of Kogut and Susskind \cite{K-S}. 

Our framework for local covariant field theory on discrete spacetimes 
is also presented in a language inspired by Oeckl's General Boundary Field Theory \cite{GBFT}. We clarify that the term used by Oeckl is General Boundary Formulation, and that it mainly concerns an axiomatic formulation of local covariant quantum physics; however, some of his more recent works deal with geometric quantization starting from classical field theories written in a form that is closely related to the one used in this article. 

A study of classical lattice gauge theory on simplicial lattices 
that allows for a Noether's theorem which is closely related to what we present in this section is presented in \cite{NoetherLGT}. 

The field equations in our framework become a set of coupled equations. 
The study of existence and uniqueness of solutions
is an issue of fundamental importance. 
Detailed studies of related existence and uniqueness problems 
have been carried on, see for example  \cite{GambiniETAL, DittrichETAL}. 

\subsection{Scalar field's action and its variation}
% - Scalar field's action and its variation
\label{DiscreteScalarGBFT}
Spacetime $M$ is given a smooth triangulation $\Delta$ (or smooth cellular decomposition of another type), and the compact connected region of our interest 
$U\subset M$ is divided into a collection of closed $n$ dimensional simplices which may intersect along their $n-1$ dimensional faces% 
\footnote{
Notice that spacetime is not required to be compact and connected; regions inside spacetime may be glued. The conditions on $U$ have the intention of making the action finite, but in many situations the conditions may be weakened. All the calculations given below will go unchanged if the variations are restricted to be of compact support. 
}
. 
A generic $n$-simplex in $U$ is denoted by 
$\nu \in U_\Delta^n$, and a generic $n-1$ simplex in $U$ is denoted by $\tau \in U_\Delta^{n-1}$. 
The boundary of our region is assumed to be an $n-1$ manifold which may have several connected components; it inherits a triangulation, and its $n-1$ simplices  will be denoted by $\tau \in (\partial U)_\Delta^{n-1}$. 

In the case of scalar fields, 
our discretization is based on decimating each $\nu \in U_\Delta^n$, keeping track only of 
one point $C\nu\in \nu^\circ$, representing ``the bulk,'' and of one point 
$C\tau\in \tau^\circ$ per boundary face, representing ``boundary face
$\tau \subset \partial \nu$.'' 
The system consists of 
a field on spacetime
while our decimated description records its value in the discrete set of points described above. 
Our record of a history is 
$\phi= \left\{ \{ \phi_\nu= \phi(C\nu)\}_{\nu \in U_\Delta^n} ,  
\{ \phi_\tau= \phi(C\tau)\}_{\tau \in U_\Delta^{n-1}} \right\}$ with all the entries in the standard fiber ${\cal F}$. 
%\begin{figure}[h!]
%\label{fig}
%  \centering
%{%
%      \includegraphics[width=0.8\textwidth]{p1.pdf}}
%%      \includegraphics[width=0.8\textwidth]{jpg1.jpg}}
%\caption{a)... b) ...}
%\end{figure}
%

\begin{figure}
\centering
\begin{subfigure}{.65\textwidth}
  \centering
  \includegraphics[width=0.8\textwidth]{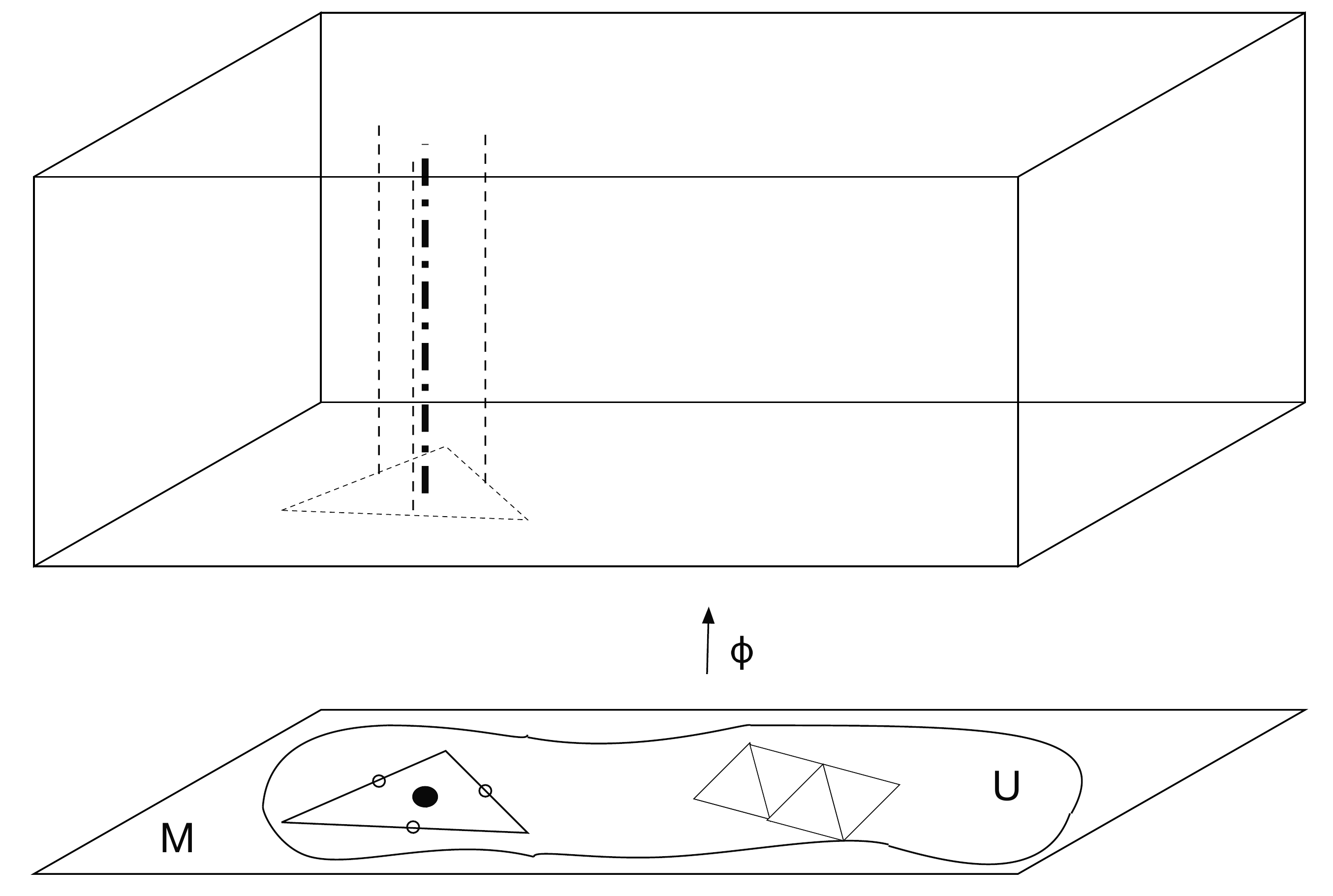}
%\label{figScalar}
\end{subfigure}%
\begin{subfigure}{.35\textwidth}
  \centering
  \includegraphics[width=0.8\textwidth]{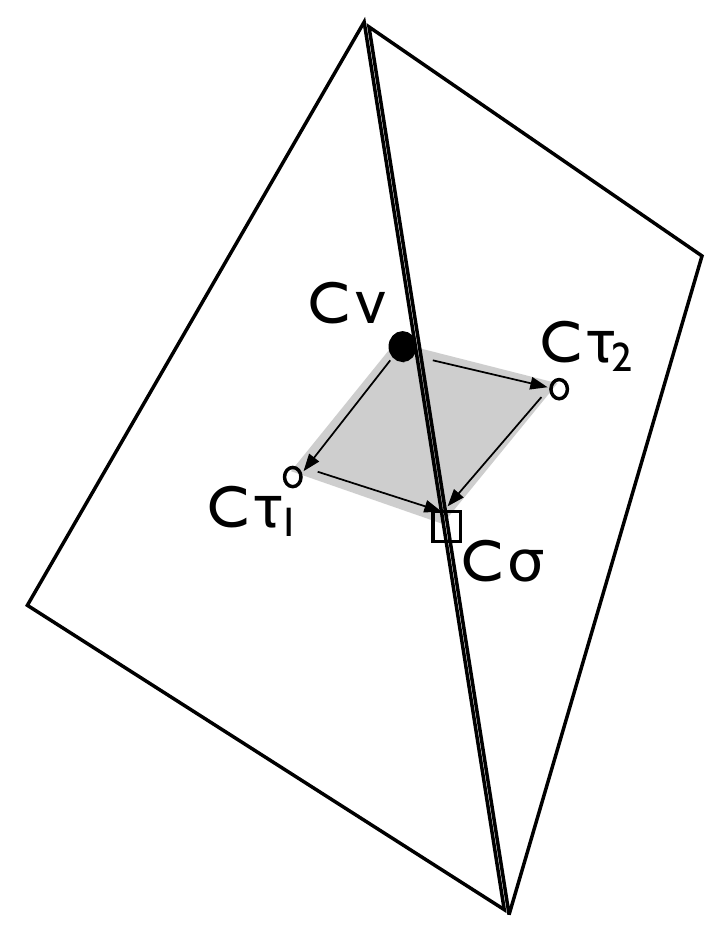}
%  \caption{A subfigure}
%  \label{figGauge}
\end{subfigure}
\caption{(a) 
A fiber bundle over a discrete set of points representing the ``bulk'' of spacetime atoms 
$\{ C\nu \}$ and the faces trough which neighboring atoms communicate 
$\{ C\tau \}$. 
(b) 
In  gauge theories local decimation stores only the parallel transport along a discrete set of paths: 
for each atom $\nu$ and each face $\tau$ in its boundary 
there is a path connecting them; additionally, for each boundary face $\tau$ and each face 
$\sigma$ in its boundary there is a path connecting them. 
}
\label{fig}
%\label{fig:test}
\end{figure}

We are interested in describing histories locally. In the continuum a history is a 
section $\phi$; 
keeping track of 
event, value of the field and partial derivatives 
$x \overset{j^1\phi}{\longmapsto}(x, \phi(x), \partial{\phi}(x))$ 
gives us the appropriate arena to study field theory in the 1st order formalism. 
Our 1st order data will be 
a decimated version of 
the portion of a history at a each atom of spacetime 
%$\nu \in U_\Delta^n$
\[
\nu \overset{\tilde{\phi}}{\longmapsto} (\nu, \phi_\nu , 
\{ \phi_\tau\}_{\tau \subset \partial \nu})
\quad . 
\]
This data gives us the value of the field $\phi_\nu=\phi(C\nu) \in {\cal F}$ at the bulk point $C\nu$, and allows us to 
estimate the first order behavior of the field in several independent ways. 
A graphical representation of the locus of these variables is shown in Figure \ref{fig}.a. 
For a moment we will imagine 
an $n$ simplex $\nu$ divided into $n+1$ corner cells $c$, one per vertex $v$ of $\nu$. A corner cell of $\nu$ 
is an $n$ cube 
with one special ``base'' vertex $C\nu$ from which $n$ edges emerge ending at 
vertices of the form $C\tau$ for the faces $\tau$ that contain $v$. 
The set of variables associated with 
each corner cell $c$ gives us 
sufficient information to give an estimate 
of the first order behavior of sections at $\nu$. 
If we had to regulate a lagrangian from the continuum we may write it in the form 
$L=\sum_c L^c$; 
in each term the value of the field would be estimated from $\phi_\nu$, and the first order behavior in $L^c$ would be estimated from the data of that corner. 
We recall that the map $\tilde{\phi}$ comes from a history, 
which implies that $\tilde{\phi}(\nu)$ and $\tilde{\phi}(\nu')$ 
are not independent if $\nu \cap \nu' = \tau \neq \emptyset$. They need to obey $\phi_\tau(\nu)= \phi_\tau(\nu')$.

Our starting point is a variational principle, and we will study the resulting geometrical structures. We start with the action 
\begin{equation}
\label{Sft}
S(\phi) = \sum_{\nu \in U_\Delta^n} 
L(\tilde{\phi}(\nu)) \quad .
\end{equation}

Hamilton's principle determines the motions predicted by our model as the extrema of the action while $\phi|_{\partial U}$ is kept fixed. 

A variation of the history, 
$\delta\phi= v= \left\{ \{ v_\nu\}_{\nu \in U_\Delta^n} ,  
\{ v_\tau\}_{\tau \in U_\Delta^{n-1}} \right\}$,  in the first order formalism is written as 
$\tilde{v}(\nu)= (v_\nu , 
\{ \tilde{\phi}(\nu)_\tau\}_{\tau \subset \partial \nu})$, where the vectors in each entry belong to $T{\cal F}$. 
Note that the relation $\phi_\tau(\nu)= \phi_\tau(\nu')$ implies 
$d\phi_\tau[\tilde{v}(\nu)]= d\phi_\tau[\tilde{v}(\nu')]$. 
The effect on the action of this variation is 
\begin{eqnarray}
\label{dSft}
dS(\phi) [v] &=& 
\sum_{\nu \in U_\Delta^n}
\frac{\partial L}{\partial \phi_\nu}(\tilde{\phi}(\nu)) d\phi_\nu [\tilde{v}(\nu)]
\nonumber\\
&+&
\sum_{\tau = \nu \cap \nu' |\nu, \nu'\in U_\Delta^n} 
%\sum_{\tau \in U_\Delta^{n-1}-(\partial U)_\Delta^{n-1}} 
\left(
\frac{\partial L}{\partial \phi_\tau}(\tilde{\phi}(\nu)) 
+ \frac{\partial L}{\partial \phi_\tau}(\tilde{\phi}(\nu')) 
\right)d\phi_\tau [\tilde{v}(\nu)]
\nonumber\\
&+& 
\sum_{\tau \in (\partial U)_\Delta^{n-1}} 
\frac{\partial L}{\partial \phi_\tau}(\tilde{\phi}(\nu_\tau)) d\phi_\tau [\tilde{v}(\nu_\tau)]
\nonumber\\
&=& 
- \sum_{U -\partial U}
\tilde{\phi}^\ast (\tilde{v} \lrcorner \hat{\Omega}_L) 
+ 
\sum_{\partial U} 
\tilde{\phi}^\ast (\tilde{v} \lrcorner \Theta_L)
\end{eqnarray}
where $\hat{\Omega}_L$ is displayed as acting on pairs composed of an $n$-chain $\tilde{\phi}(\nu)$ and a vector 
$\tilde{v}(\nu)=(v_\nu , \{ v_\tau\}_{\tau \subset \partial \nu})$,  
while $\Theta_L$ 
acts on pairs composed of an $n-1$-chain $\tilde{\phi}(\tau)$ and a vector 
corresponding to the variation of the history at $\nu(\tau)$. 
Our definitions of the structural forms are 
\[
\Theta_L(\cdot , \tilde{\phi}(\tau_\nu)) = 
%sgn(\nu , \tau) 
\frac{\partial L}{\partial \phi_\tau}(\tilde{\phi}(\nu)) d\phi_\tau
\quad , 
\]
\[
\hat{\Omega}_L(\cdot , \tilde{\phi}(\nu)) = 
- \frac{\partial L}{\partial \phi_\nu}(\tilde{\phi}(\nu)) d\phi_\nu - 
\sum_{\tau \in (\partial \nu)^{n-1}}
%sgn(\nu , \tau) 
\frac{\partial L}{\partial \phi_\tau}(\tilde{\phi}(\nu)) d\phi_\tau
\quad . 
\]
The form $\hat{\Omega}_L(\cdot , \tilde{\phi}(\nu))$ is seen as having a term corresponding to the bulk of 
$\tilde{\phi}(\nu)$ and one term per $n-1$ face of $\partial\nu$. In the last equality of formula (\ref{dSft}) we have used this decomposition of the terms in $\hat{\Omega}_L$, and 
the first sum excludes the terms corresponding to $n-1$ faces in $\partial U$. 
Also we clarify that in Equation (\ref{dSft}) the argument of Cartan's form $\Theta_L$
includes $\tilde{\phi}(\tau_\nu)$ with 
the $n$ simplex satisfying $\tau \subset \nu \subset U$. 
This notation for the variation of the action 
will be very convenient; however, we warn the reader that 
the summation over $U -\partial U$ defined above is not a standard operation among 
cochains. 
In Section \ref{1d} we gave a detailed explanation of 
the notation used above to write the variation of the action after Equation (\ref{dS}) . 

Returning to Hamilton's principle, from the terms in Equation (\ref{dSft}) 
corresponding to the interior of $U$ 
(information also stored in $\hat{\Omega}_L$) 
we read two types of field equations: (i) 
$\frac{\partial L}{\partial \phi_\nu}(\tilde{\phi}(\nu))=0$, equations ensuring that we have a solution in the interior of each $n$ simplex $\nu$. (ii) 
$\frac{\partial L}{\partial \phi_\tau}(\tilde{\phi}(\nu)) 
+ \frac{\partial L}{\partial \phi_\tau}(\tilde{\phi}(\nu')) =0$ with $\nu \cap \nu' = \tau$, conditions that paste local solutions involving 
data on both sides of the $n-1$ simplex $\tau$ in the interior of $U$.

\subsection{Gauge field's action and its variation}
% - Gauge field's action and its variation
\label{gauge}
Our notation regarding the discretization of spacetime $M$ will follow closely the terminology of the previous subsection. One difference is 
that now we will need to refer to $n-2$ simplices which will be denoted generically by the letter $\sigma$, and the chosen point in the interior of $\sigma$ will be denoted by $C\sigma$. A deeper difference is that gauge fields are not decimated at points like we did with scalar fields; instead lattice gauge theories decimate by ignoring the connection everywhere except for what is necessary to calculate the parallel transport along a preferred set of oriented links. The variables in this framework are local in the sense that each variable is related to a single $n$ simplex $\nu$, with some identities between variables related to links shared by two $n$ simplices. The variables of this framework were introduced by Reisenberger \cite{Reisenberger}, and our notation is compatible with his. The variables are 
\[
\{ h_l \in G \} \quad , \quad \{ k_r \in G \} ,
\]
where each link $l$ lies in the interior of an $n$ simplex $\nu$ going from $C\nu$ to a $C\tau$, and each link $r$ lies in an $n-1$ simplex $\tau$  
going from $C\tau$ to a $C\sigma$; 
see Figure \ref{fig}.b. 
In an $n$ simplex
there are $n+1$ interior links and $(n+1)n$ links that lie in the boundary. 
After reference points in the fiber over the points $(C\nu, \{C\tau \}, \{ C\sigma \})$ are fixed, we can parametrize parallel transport along the links by group elements; this is the meaning that we assign to the variables. 
If the connection changes by an internal gauge transformation, our decimated history changes. The group $G$ acts independently on the fibers over $(C\nu, \{C\tau \}, \{ C\sigma \})$ and changes the parallel transport variables by conjugation. 
We remark that in the calculations that we will present below we treat the case in which the Lie group $G$ is a matrix group. 

The smallest circuits that can be made with these links 
and their associated group element 
will play an important role; we will write $g_{\partial s} = h_{l2}^{-1} k_{r2}^{-1} k_{r1} h_{l1}$, where 
a {\em wedge} $s$ is an oriented $2$-chain inside an $n$ simplex $\nu$ (the shaded region in Figure \ref{fig}.b) 
whose boundary links are those indicated in the formula for $g_{\partial s}$. 
Clearly, the orientation of $s$ must be prescribed before $\partial s$ makes sense as an oriented path. We will leave the orientation free and keep in mind that if we write $\bar{s}$ for the same wedge but with the opposite orientation, then
$g_{\partial \bar{s}}= g_{\partial s}^{-1}$. 

Our record of a history is 
$A= \left\{ \{ h_l \} ,  \{ k_r \} \right\}_U$, where the collection includes all the links of types $l$ and $r$ in the discretization of $U$, and all the entries belong to $G$. 

The local first order data in this discrete framework is 
a decimated version of 
the portion of a history at each atom of spacetime $\nu \in U_\Delta^n$, 
\[
\nu \overset{\tilde{A}}{\longmapsto} 
(\nu, \{ h_l \}_\nu , \{ k_r \}_\nu)
\quad . 
\]
The space of local first order data plays a central role because it 
is the domain of the discrete lagrangian density; for a graphical representation of the locus of these local degrees of freedom see Figure \ref{fig}.b. Apart from one discrete label, the rest is composed of several copies of the matrix group $G$. In calculations, our notation makes use of the fact that, apart from the discrete label, the domain is a subspace of some $\R^N$. 
An interpretation of this data set is that the $h$ matrices have decimated information regarding the connection $1$-form 
inside $\nu$, and the data set also 
contains decimated information regarding the connection's curvature in $\nu$. 
Our knowledge regarding curvature of the connection at $\nu$ is contained in the collection 
\[
\{ g_{\partial s}\}_\nu  \quad . 
\]
Each corner cell $c$ in $\nu$ gives us a minimal set of variables that permits us to 
estimate the curvature of the connection in this discrete framework. 
Thus, 
if we had to regulate a lagrangian from the continuum we may write it in the form $L=\sum_c L^c$. 
We recall that the map $\tilde{A}$ comes from a history, 
which implies that $\tilde{A}(\nu)$ and $\tilde{A}(\nu')$ 
are not independent if $\nu \cap \nu' = \tau$ is an $n-1$ simplex of $\Delta$. They need to 
agree on $k_r$ for all links $r$ in $\tau$. 

Again, we start from Hamilton's principle, and 
our study will focus on the resulting geometrical structures. 
The most natural actions within this framework% 
\footnote{
Actions that are one step less local will be considered as reduced models in Subsection 
\ref{lgt}; those actions are closely related to Wilson's action on ordinary lattices. 
} 
are of the type 
\begin{equation}
\label{Sgauge}
S(A) = \sum_{\nu \in U_\Delta^n} L(\tilde{A}(\nu)) 
= \sum_{\nu \in U_\Delta^n} L(\nu, \{ g_{\partial s}\}) \quad .
\end{equation}
They are as local as possible, 
and gauge invariance needs to be analyzed only at the points $C\nu$.
An explicit formula for $L(\nu, \{ g_{\partial s}\})$ requires a regularization; see Section \ref{Reg-Coar-Cont}. 

We recall that at each atom of spacetime $\nu$ we can separate bulk degrees of freedom from boundary degrees of freedom: the variables $h_l$ for $l\subset \nu$ describe bulk degrees of freedom, while variables $k_r$ for $r\subset \partial \nu$ describe degrees of freedom associated with the boundary face $\tau$ that contains $r$. When we consider a region composed of many atoms, the boundary of the region will be decomposed in $n-1$ simplices and the boundary degrees of freedom will be described by variables $k_r$ for $r\subset \partial U$. 
This ability to cleanly factorize the degrees of freedom into bulk and boundary degrees of freedom is a key feature to identify the geometrical structure that arises from the variations of the action. 

Since $G$ is a matrix group, in a variation of a history each entry is also a matrix, 
$\delta A= v= \left\{ \{ v_l \in T_{h_l}G\} ,  \{ v_r  \in T_{k_r}G\} \right\}_U$. 
We will later parametrize such variations by left- or right-invariant vector fields at our convenience, but at the beginning stage we will keep the notation open. 
In the first order formalism we will write 
$\tilde{v}(\nu)=(\{ v_l \}_\nu , \{ v_r \}_\nu)$. 
In Subsection \ref{1dExamples} we presented two models describing 
rigid body motion, which use group variables; 
reviewing these discrete time models might be insightful. 

The effect of a variation on the action is 
\begin{eqnarray}
\label{dSgauge}
dS(A) [v] &=& 
\sum_{\overset{\nu \in U_\Delta^n}{l \subset \nu}}
\frac{\partial L}{\partial h_l}(\tilde{A}(\nu)) d h_l [\tilde{v}(\nu)]
\nonumber\\
&+& 
\sum_{\overset{\tau = \nu \cap \nu' |\nu, \nu'\in U_\Delta^n}
{r \subset \tau}} 
\left(
\frac{\partial L}{\partial k_r}(\tilde{A}(\nu)) 
+ \frac{\partial L}{\partial k_r}(\tilde{A}(\nu'))
\right) d k_r [\tilde{v}(\nu)]
\nonumber\\
&+& 
\sum_{\overset{\tau \in (\partial U)_\Delta^{n-1}}{r \subset \tau}} 
\frac{\partial L}{\partial k_r}(\tilde{A}(\nu_\tau)) d k_r [\tilde{v}(\nu_\tau)]
\nonumber\\
&=& 
- \sum_{U -\partial U}
\tilde{A}^\ast(
\tilde{v} \lrcorner \hat{\Omega}_L) 
+ 
\sum_{\partial U} 
\tilde{A}^\ast(
\tilde{v} \lrcorner \Theta_L) . 
\end{eqnarray}

As in the case of scalar fields, $\hat{\Omega}_L$ acts 
on pairs composed of 
a vector $\tilde{v}(\nu)$ and 
an $n$-chain $\tilde{A}(\nu)$,  
while $\Theta_L$ 
acts on pairs composed of 
the variation of the history at $\tau$ and 
an $n-1$-chain $\tilde{A}(\tau)$. 
The resulting geometric structure is, then, 
\[
\Theta_L(\cdot , \tilde{A}(\tau_\nu)) = 
\sum_{r \subset \tau} 
\frac{\partial L}{\partial k_r}(\tilde{A}(\nu(\tau))) d k_r
\quad , 
\]
\[
\hat{\Omega}_L(\cdot , \tilde{A}(\nu)) = 
- \sum_{l \subset \nu}
\frac{\partial L}{\partial h_l}(\tilde{A}(\nu)) d h_l - 
\sum_{r \subset \nu} 
\frac{\partial L}{\partial k_r}(\tilde{A}(\nu)) d k_r 
\quad . 
\]
Writing 
Cartan's form as 
$\Theta_L(\tilde{v}(\nu) , \tilde{A}(\tau_\nu)) = 
(\tilde{v}(\nu) L) (\tilde{A}(\nu))|_\tau$ will be useful: the contribution from the $\tau$ degrees of freedom to 
the derivative operator $\tilde{v}(\nu)$ acting on $L$ evaluated at $\tilde{A}(\nu)$. 

Let us briefly comment on the field equations resulting from (\ref{dSgauge}) for 
any model of this type; later we will study them with more detail within specific examples. 
Viewing the field equations as divided into two types is insightful: 
(i) A set of field equations of the type 
$\{ \frac{\partial L}{\partial h_l}(\tilde{A}(\nu))=0 \}_{l\subset \nu}$
requires that the history be 
a solution in the interior of the atom $\nu$. 
(ii) A set of equations of the type 
$\{\frac{\partial L}{\partial k_r}(\tilde{A}(\nu))
+ \frac{\partial L}{\partial k_r}(\tilde{A}(\nu'))=0 \}_{r\subset \tau}$ (with 
$\nu \cap \nu' = \tau$) 
gives gluing conditions for the solutions on the two atoms 
$\nu, \nu'\subset U$ that meet at face $\tau$.

Internal gauge transformations induce specific types of variations. A decimated history over an atom $\nu$ is sensitive to internal gauge transformations on the fibers over points the $(C\nu, \{C\tau \}, \{ C\sigma \})$. If our lagrangian is of the form $L(\nu, \{ g_{\partial s}\})$ and it is 
invariant under gauge transformations over $C\nu$,  
then the variation of the action caused by any internal gauge transformation vanishes for every history. 
This means that the system of field equations is redundant. 
A lagrangian of the type we are considering would allow us to work at the level of gauge equivalence classes, but we will not do that in this article. 
In Subsection 
\ref{Noether} we will see that there are local constraints on the boundary data associated with the invariance under internal gauge transformations.

\subsection{Actions of BF type and their variation}
% - Actions of BF type and their variation
\label{BF+subsection}
Now we consider field theories with degrees of freedom captured by 
a connection, a Lie algebra valued $n-2$ form and a spacetime scalar object that is used as a Lagrange multiplier, $(A, B, \varphi)$. 
In the previous two subsections we described how we treat connections and spacetime scalars. Since the partial derivatives of the spacetime scalar do not appear in the lagrangian, the treatment of $\varphi$ is simplified, and we decimate the field only at the center of the atoms, $\varphi_\nu=\varphi(C\nu)$. The decimation of the Lie algebra valued $n-2$ form needs to be done in such a way that the result can be appropriately coupled to the discrete curvature $g_{\partial s}$. In addition, since partial derivatives of this field do not enter in the lagrangian it would be convenient to assign them a discrete counterpart that is ``internal'' to the atoms. Reisenberger's discretization/decimation  meets all these requirements by assigning a variable $e_s\in Lie(G)$ to each wedge $s$. The decimation formula (see Section 4 of \cite{Reisenberger}) involves an integral over the co-dimension two simplex in $\nu$ dual to $s$, but 
the formula makes $e_s$ sensitive only to gauge transformations over $C\nu$. 
Under a change of orientation of the wedge, the variable changes sign:  
$e_{{s'}}=- e_s$. 
For examples of discrete theories using this variable see \cite{Reisenberger, Reisenberger94, F-K}. 
Our geometric structures follow from treating this variable as describing degrees of freedom  ``internal'' to $\nu$.  

The record of a history is 
$\ae= \left\{ \{ h_l \} ,  \{ k_r \} , \{ e_s \}, \{ \varphi_\nu \} \right\}_U$, where the collection includes variables associated with all the links of types $l$ and $r$, all wedges $s$ and atoms $\nu$ inside $U$. 

The local first order data in this discrete framework is 
a decimated version of 
the portion of a history at a each atom of spacetime $\nu \in U_\Delta^n$
\[
\nu \overset{\tilde{\ae}}{\longmapsto} 
(\nu, \{ h_l \}_\nu , \{ k_r \}_\nu , \{ e_s \}_\nu , \varphi_\nu )
\quad . 
\]
Again we remark that the map $\tilde{\ae}$ comes from a history, 
which implies that $\tilde{\ae}(\nu)$ and $\tilde{\ae}(\nu')$ 
are not independent if $\nu \cap \nu' = \tau$ is an $n-1$ simplex of $\Delta$. 
In this case the shared boundary data is purely connection; 
the only conditions are $k_r(\tilde{\ae}(\nu))= k_r(\tilde{\ae}(\nu'))$ for all 
links $r$ in $\tau$. 

The actions that we consider are of the type 
\begin{equation}
\label{SBF+}
S(\ae) = \sum_{\nu \in U_\Delta^n}
\left[ 
\sum_{s \subset \nu} e_s^i \theta_{s, i} + \Phi(\nu, \{ e_s \}, \varphi) 
\right] \quad , 
\end{equation}
where $\theta_{s, i}$ has two alternative definitions: (i) 
$\theta_{s, i} = \Tr(f_i^Tg_{\partial s})$ if $G$ is a group of real matrices, and $\{ f_i \}$ is an orthonormal basis of $Lie (G)$ according to the matrix inner product $(A, B) = \Tr(A^T B)$. 
(ii) $\theta_{s, i} = \Rp\Tr(f_i^\dagger g_{\partial s})$ if $G$ is a group of complex matrices, and $\{ f_i \}$ is an orthonormal basis according to $(A, B) = \Tr(A^\dagger B)$. 
Notice that under a change of orientation of $s$ the term $e_s^i \theta_{s, i}$ remains invariant. 
In specific examples we may change the conventions by rescaling the inner product and the generators appropriately. 
The variable $\varphi \in V$ belongs to some vector space and appears as a Lagrange multiplier which will induce a constraint on the variables $e_s$. 

We will study the variation of this type of action following the same steps as we did in the previous subsection. 
As a preamble to this type of model, in Subsection \ref{1dExamples}, we presented a model of rigid body motion with discrete time which uses group variables and Lie algebra variables while mirroring the mathematical structure of the family of discrete spacetime models that we now study. 

Our notation for variations in the local 1st order format is 
$\delta \tilde{\ae}(\nu)=\tilde{v}(\nu)=(\{ v_l \in T_{h_l}G \}_\nu , 
\{ v_r \in T_{k_r}G \}_\nu , \{ v_s \in T_{id}G \}_\nu , v_\nu \in V)$. 
Now separate bulk degrees of freedom from boundary degrees of freedom at each atom $\nu$: the variables $h_l$ for all $l\subset \nu$ describe bulk degrees of freedom, 
as well as the variables $e_s$ for all $s\subset \nu$ and the variable $\varphi_\nu$, 
while variables $k_r$ for all $r\subset \tau \subset \partial \nu$ describe degrees of freedom associated with boundary face $\tau$. An important feature of this family of models is that the boundary degrees of freedom of each atom, $\nu$, are purely connection. 

The variation of the action can be written as 
\[
dS(\ae) [v] =
- \sum_{U -\partial U}
\tilde{\ae}^\ast(
\tilde{v} \lrcorner \hat{\Omega}_L) 
+ 
\sum_{\partial U} 
\tilde{\ae}^\ast(
\tilde{v} \lrcorner \Theta_L)
\]
with 
\[
\Theta_L(\cdot , \tilde{\ae}(\tau_\nu)) = 
\sum_{r \subset \tau} 
\frac{\partial L}{\partial k_r}(\tilde{\ae}(\nu(\tau))) d k_r
\quad , 
\]
\begin{eqnarray*}
\hat{\Omega}_L(\cdot , \tilde{\ae}(\nu)) =&& 
- \sum_{l \subset \nu}
\frac{\partial L}{\partial h_l}(\tilde{\ae}(\nu)) d h_l - 
\sum_{r \subset \nu} 
\frac{\partial L}{\partial k_r}(\tilde{\ae}(\nu)) d k_r \\
&&
-\frac{\partial \Phi}{\partial \varphi}(\tilde{\ae}(\nu)) d \varphi
- \sum_{l \subset \nu}
\frac{\partial L}{\partial e_s}(\tilde{\ae}(\nu)) d e_s
\quad . 
\end{eqnarray*}

There are a few general features of the field equations of models in this family: 
Since $\Phi$ depends only on internal degrees of freedom for every $\nu$, 
the gluing field equations will be exactly the same for all models of the family. Furthermore, the field equations related to the variation of the $h_l$ variables will be the same for all the models in this family. 
Also, the discussion regarding gauge symmetries given in the previous subsection applies to models of this family without any change. Gauge symmetries of the action imply that the system of field equations is redundant, and in Subsection \ref{Noether} we will see that there are local constraints on the boundary data associated with invariance under internal gauge transformations. 

In Subsection \ref{MichaelsModel} we briefly comment on specific features of the field equation of Reisenberger's model; for a more detailed study of the model see \cite{Reisenberger}.

\subsection{The multisymplectic formula}
% - The multisymplectic formula
\label{MutisymplFormulaSect}
We will define the multisymplectic form $\Omega_L$ 
acting on two vectors induced by 
``vertical'' variations of the history. 
First we will introduce this concept for the {\em scalar field}. 
Consider two variations in the first order format $\tilde{v}, \tilde{w}$; we then define 
\[
\Omega_L ( \tilde{v}, \tilde{w} , \tilde{\phi}(\tau_\nu)) = 
- d(\Theta_L|_{\tilde{\phi}(\tau_\nu)}) (\tilde{v}, \tilde{w}) . 
\] 
It is natural to consider discrete lagrangians for which 
$\frac{\partial L}{\partial \phi_\tau}(\tilde{\phi}(\nu))$ is a function of 
$\phi_\nu , \phi_\tau$. For lagrangians of this type we get 
\[
\Omega_L ( \cdot, \cdot , \tilde{\phi}(\tau_\nu)) = 
-\frac{\partial^2 L}{\partial \phi_\nu \partial \phi_\tau}(\tilde{\phi}(\nu)) 
d\phi_\nu \wedge d\phi_\tau . 
\]

The objects defined now and in previous subsections 
acquire useful properties when only solutions and first variations 
are considered. 
In this context, according to (\ref{dSft}), we can write 
$dS(\phi) [v] = \sum_{\partial U} \tilde{\phi}^\ast(
\tilde{v} \lrcorner \Theta_L)$. 
Therefore, when imported to the space of solutions, the following integral of the 
multisymplectic form vanishes, 
$\sum_{\partial U} \tilde{\phi}^\ast \Omega_L = -ddS=0$. More precisely, 
\[
\sum_{\partial U} 
\tilde{\phi}^\ast(
\tilde{w} \lrcorner \tilde{v} \lrcorner
\Omega_L) =0 
\]
for any first variations $v, w$ of any solution of the field equations $\phi$ in $U$. 
This is the multisymplectic formula 
in this discrete version of multisymplectic covariant field theory. 

We will study the meaning of the multisymplectic formula using GBFT language. 
%-dimensional
Consider the $n$-dimensional region $U$ whose boundary may have several connected components, $\partial U = B_1 \cup ... \cup B_m$. At the first stage of this study we will isolate each component from the rest. Each $B_i$ has no boundary, and we will consider the smallest neighborhood of $B_i$ inside $U$. We will consider the case in which there is a region in our triangulation $U(B_i)\subset U$ homeomorphic to $B_i \times [0, 1]$ and with 
$\partial U(B_i)= B_i - B'_i$; the interpretation is that a homotopy has been used to move $B_i$ by one step 
inside $U$ without leaving any point fixed, and we have obtained $B'_i$. 
We furthermore restrict the study to the case in which $U(B_i) \cap U(B_j)= \emptyset$ 
whenever $i\neq j$. 
The formalism described above can be applied to each $U(B_i)$ independently. The result is a space $\mbox{Sols}(U(B_i))$ of solutions (with its corresponding space of first variations) equipped with a nontrivial $2$-form $\omega_{L,i}=-d\theta_{L,i}$ that may be evaluated from 
$\Omega_L=-d\Theta_L$ used as a cochain on 
$B_i$ (or using $\Omega_L$ on $B'_i$ yielding the same result due to the multisymplectic formula for $U(B_i)$). 
Our assumptions imply that in the multicomponent region 
$U(\partial U)=\cup_i U(B_i)$ the space of solutions has the nontrivial structure 
$\mbox{diag}(\omega_{L, 1}, ... , \omega_{L, m})$. 
In this setting we can see how 
solutions in the whole $U$ correlate the spaces 
$\mbox{Sols}(U(B_i))$; we will describe this situation using the map 
$I_U:\mbox{Sols}(U) \to \mbox{Sols}(U(\partial U)) = \times_i \mbox{Sols}(U(B_i))$ 
that restricts solutions in $U$ to $U(\partial U)=\cup_i U(B_i)$. 
The multisymplectic formula implies that 
\[
I_U^\ast \mbox{diag}(\omega_{L, 1}, ... , 
\omega_{L, m}) = 0 \quad . 
\]
Each component of $\partial U$ contributes a nontrivial term to the equation, while the field equations intervene to correlate all the terms providing a conservation law. 
Symplectic geometry reads the previous equation as solutions in $U$ inducing isotropic submanifolds of the boundary phase space; 
lagrangian submanifolds in classical GBFT in the continuum play an important role \cite{GBFT}. 

An important application of the previous formula in a region $U$ of the form 
$\Sigma \times [0, 1]$ implies that solutions of the field equations ``transport'' first variations in a way that preserves the symplectic form.

Now we will address the case of {\em gauge fields}; we will arrive at a multisymplectic formula in that setting. 
The 
multisymplectic form $\Omega_L$ 
acting on two variations in first order format 
$\tilde{v}, \tilde{w}$ of the history at the boundary is 
\[
\Omega_L ( \tilde{v}, \tilde{w} , \tilde{A}(\tau_\nu)) = 
- d(\Theta_L|_{\tilde{A}(\tau_\nu)}) (\tilde{v}, \tilde{w}) . 
\] 
In Section \ref{FTexamples} we will 
study lagrangians of the form 
$L(\tilde{A}(\nu))= \sum_s L^s(g_{\partial s})$. 
In particular, we will focus on a regularization of pure Yang-Mills theory on regular lattices and write the multisymplectic formula explicitly.

Let us now consider only histories that are solutions and variations that are first variations; in this case, (\ref{dSgauge}) implies 
$dS(A) [v] = \sum_{\partial U} \tilde{A}^\ast(\tilde{v} \lrcorner \Theta_L)$. 
Thus, when imported to the space of solutions the following integral of the 
multisymplectic form vanishes, 
$\sum_{\partial U} \tilde{A}^\ast \Omega_L = -ddS=0$. More precisely, 
for any first variations $v, w$ of any solution of the field equations $A$ we have 
\[
\sum_{\partial U} \tilde{A}^\ast(\tilde{w} \lrcorner \tilde{v} \lrcorner
\Omega_L) =0 . 
\]
This is the multisymplectic formula 
in this framework for gauge theory on discrete spacetimes. 

A study of the meaning of the multisymplectic formula using GBFT language follows the same logical path as the case of scalar field; we will not repeat it.

\subsection{Symmetries and conserved quantities}
% - Symmetries and conserved quantities 
\label{Noether}
We will study first the {\em scalar field}. 
Let a Lie group ${\cal G}$ act on the standard fiber 
${\cal F}$ and on histories in a diagonal form: the action on histories is 
$\phi= \left\{ \{ \phi_\nu\}_{\nu \in U_\Delta^n} ,  
\{ \phi_\tau\}_{\tau \in U_\Delta^{n-1}} \right\} \overset{g}{\longmapsto} 
\left\{ \{ g(\phi_\nu)\}_{\nu \in U_\Delta^n} ,  
\{ g(\phi_\tau)\}_{\tau \in U_\Delta^{n-1}} \right\}$, and for 
histories in the local 
first order format we have 
$(g\tilde{\phi})(\nu) = (\nu, g(\phi_\nu) , \{ g(\phi_\tau)\}_{\tau \subset \partial \nu})$. 

Now consider the case in which this action  
leaves the discrete lagrangian invariant: 
$L(g\tilde{\phi}(\nu)) = L(\tilde{\phi}(\nu))$ for all $\nu \in U_{\tiny\mbox{disc}}$, for any first order history 
$\tilde{\phi}$ and for any $g\in {\cal G}$. 
In this case $S$ is also invariant, and the ${\cal G}$-action preserves the subspace of extrema.
The fact that the subspace of extrema has null directions of $dS$ leads to conserved quantities. Below we state this version of Noether's theorem in this framework of field theory in discrete spacetime. 

The first variation induced by the infinitesimal 
${\cal G}$-action in the direction of $\xi\in Lie({\cal G})$ is a vector field that we denote by 
$v_\xi$. Given any solution $\phi$ we have 
\[
0 = dS(\phi) [v_\xi] = 
\sum_{\partial U} \tilde{\phi}^\ast(
\tilde{v}_\xi \lrcorner \Theta_L)
 . 
\]
This is the expression of Noether's theorem in this discrete spacetime mutisymplectic formalism. 

In GBFT language, using the objects introduced in the previous subsection, 
the result is stated as the conservation law 
\[
\sum_{i=1}^m J_\xi(B_i)=0 \, , \mbox{ where } \, \,
J_\xi(B_i)= \sum_{B_i} \tilde{\phi}^\ast(
\tilde{v}_\xi \lrcorner \Theta_L)= \theta_{L, i}(v_\xi|_{B_i}) . 
\]

% FINISHES scalar field case

In the case of {\em gauge fields} 
the most prominent symmetry transformations are 
internal gauge transformations. 
We previously mentioned that an internal gauge transformation modifies the decimated history in a very simple form: 
the structure group $G$ acts {\em independently} at each fiber over the points of our discretization. The discrete lagrangians that we consider are invariant under this group of transformations. 
Thus, the formalism will yield ``conserved quantities'' only in the sense that an equation of the form $0=\sum_{\partial U} \tilde{A}^\ast(\tilde{v}_\xi \lrcorner \Theta_L)$ holds 
when evaluated on solutions and for all first variations $v_\xi$ associated with our symmetry. However, 
we will see below that a more careful analysis shows that 
these equations can also be seen as producing redundancies among the field equations or as local constraints. 

First consider a single spacetime atom $\nu$. Let $A$ be a solution on $\nu$. 
We start with the study of the effects of a first variation induced by an internal gauge transformation on the fiber over $C\nu$. 
This variation vanishes in the boundary degrees of freedom; thus, it does not yield any conserved quantity or constraint; instead, it contributes to the redundancies of the field equations. 

We continue considering a single spacetime atom $\nu$, but let us turn our attention to 
internal gauge transformations over points of the type $C\tau$. 
Recall that there is only one link $l$ in $\nu$ touching $C\tau$, 
and there are $n$ links $r$ starting from $C\tau$. 
This type of first variation has an effect on the bulk variable $h_l$ and on the boundary variables $\{ k_r \}_\tau$. 
A first variation induced by the infinitesimal internal gauge transformation over $C\tau$ 
may be parametrized by $\xi \in \mbox{Lie}(G)=T_{\mbox{id}}G$. We remind the reader that 
$G$ is a matrix group, which permits us to treat Lie algebra elements also as matrices. 
The effect of the variation on $h_l$ is the left-translation $\delta h_l= \xi h_l \in T_{h_l}G$, and its effect on a variable $k_r$ is the right-translation 
$\delta k_r= k_r \xi \in T_{k_r}G$ (one variation for each of the links of type $r$ starting at $C\tau$, all induced by the same $\xi$). 
Since the variable $h_l$ is considered internal to $\nu$, while the other variables involved --the $k_r$ variables-- are not internal to $\nu$, 
the field equations of the interior of $\nu$ only imply that the contribution of $\delta h_l$ vanishes leading to 
$0=dS(A) [v_\xi] = \sum_{\partial \nu} \Theta_L(\delta \tilde{v}_\xi , \tilde{A}(\tau_\nu)) = 
\sum_{r \subset \tau} 
\frac{\partial L}{\partial k_r}(\tilde{A}(\nu(\tau))) (k_r \xi)$
{\em for all} $\xi \in \mbox{Lie}(G)$. 
If the given relation does not hold, there cannot be a solution compatible with the boundary data; thus, we have found a {\em local constraint} on the boundary data. We will see the explicit form of this constraint in concrete examples. 

Now let us study first variations due to 
internal gauge transformations over points of the type $C\sigma$. 
In $\nu$ there are two links $r1\subset \tau 1$ and $r2\subset \tau 2$ 
finishing at $C\sigma$; therefore, there are two terms in the equation 
$0=dS(A) [v_\xi]$, and 
field equations are not involved. 
This expression holds for every history, and it cannot be simplified/modified by field equations. The dynamical content of the boundary data on $\partial \nu$ is completely captured by the reduced variable $k_{r2}^{-1}k_{r1}$; however, we do not 
eliminate the rest of the boundary variables 
because pasting $\nu$ to other atoms would be impossible.

What happens with this study of conserved quantities from internal gauge transformations when we study several atoms amalgamated to form a spacetime region $U\subset M$? 
One needs to untangle bulk degrees of freedom, which must obey field equations, from boundary degrees of freedom, and then study the invariances of the action. The results are as follows: (i) invariance of the action under internal gauge transformations on fibers over bulk points contributes to the redundancy of the field equations, and (ii) invariance under internal gauge transformations on fibers over boundary points ($C\tau$ or $C\sigma$) where boundary links and bulk links meet yields local constraints on boundary data.

% finishes 

The action under study may be invariant under more involved transformations, 
transformations that may involve several neighboring spacetime atoms or completely nonlocal transformations. 
As long as the transformations preserve the action, 
they also preserve the subspace of extrema, and the ideas used in the previous 
study apply to them. Using our geometric tools one would find explicit expressions for 
conserved quantities and/or local constraints, depending on the specific case.

\section{Atomic boundary data formalism}
\label{BdaryDataF}
%\subsection{Boundary data formalism}

We may be especially interested in 
a model for a field theory on a discrete spacetime 
in which solutions are composed by gluing histories over spacetime atoms that are not divided into corners. 
Neighboring atoms would share boundary data, and the gluing conditions would be the only field equations. 

The first option to try is to define a reduced system solving the internal field equations at each $\nu$. The reduced lagrangian is 
$L^b(\tilde{\phi}^b(\nu))= L(\tilde{\phi}(\nu))$ for 
$\tilde{\phi}(\nu)$ an extremum of $L$ for fixed boundary conditions $\tilde{\phi}^b(\nu)$. 
The reduced system has field equations determined by $L^b$, and the new version of Cartan's form $\Theta_L^b$ is induced: 
\[
\Theta_{L^b}(\cdot , \tilde{\phi}^b(\tau_\nu)) = 
%sgn(\nu , \tau) 
\frac{\partial L^b}{\partial \phi_\tau}(\tilde{\phi}^b(\nu)) d\phi^b_\tau .  
\]
There is then a multisymplectic form $\Omega_{L^b}=-d\Theta_{L^b}$ that participates in the appropriate multisymplectic formula. 
Furthermore, solving the gluing field equations is equivalent to 
solving the implicit equations defined by the coupled 
covariant local generating functions $L^b$ corresponding to the neighboring atoms. 

A geometrical interpretation of this ``evolution by canonical moves'' in terms of Hamilton's principal function is sketched in the opening paragraphs of Section \ref{1d}. We remark that here this picture is merely motivational because in effective theories for nontrivial 
field theories on spacetimes of dimension higher than $1$ we do not have access to Hamilton's function. For an extended discussion see Section \ref{Reg-Coar-Cont}.

Clearly the functional form of $L^b$ is in general complicated since it requires solving nontrivial field equations of the nonreduced system. In the following subsection we describe a different reduction which leads to a simple lagrangian. 

A second option to try is defining a simple effective theory directly on a discretization in which the only data stored for a given spacetime atom $\nu$ is the data over 
$\partial \nu$.  

Notice that a discrete field theory of this type is not defined on an ordinary lattice. For example, in the case of scalar fields the lattice sites where degrees of freedom sit are the points of type $C\tau$ that are shared by $2$ atoms independently of the dimension; in contrast, for an ordinary cubical lattice 
each site has $2n$ neighbors.

\section{Reduced formalism}
\label{Reduced}
%\label{Reduced1d}

We tried to emphasize that our formalism had two types of field equations: (i) equations in the interior of the atoms and (ii) simple gluing equations. 
Since the gluing equations are simple, we can solve them explicitly and end up with a formalism with fewer variables, the ``bulk variables,'' and a reduced lagrangian that is not too complicated. 

Histories in this reduced formalism, $\phi^r=\{ \phi_\nu\}_{\nu \subset U}$ in the case of the scalar field, are histories in an ordinary lattice discretization of spacetime 
where all the sites are of the same type. 
The values of the rest of the variables of our non-reduced model will be inferred by explicitly solving the gluing equations $\phi=\phi(\phi^r)$. Nontrivial field equations are 
determined by the reduced action $S_r(\phi^r)= \sum_{Ur} L(\widetilde{\phi(\phi^r)}(\nu))$ that is defined on a slightly contracted domain $U_r \subset U$ which has points of the type $C\nu$ in its boundary. 

The reduced action can be written in terms of a reduced lagrangian 
$S_r(\phi^r)= \sum_{Ur} L^r(\tilde{\phi^r})$, where the sum is over cells dual to vertices of our triangulation. The field restricted to a dual cell $\Box$, 
$\phi^r=\{ \phi_\nu\}_{C\nu \in \Box}$, is the argument of the reduced lagrangian. 
If the original lagrangian was written as a sum of corner lagrangians $L= \sum_c L^c$, 
the reduced lagrangian is also a sum of corner lagrangians 
$L^r=\sum_{c\subset \Box} L^c$. 
We show explicitly how this reduction is done for scalar fields in Subsection \ref{nlwaves}.

If solutions of the gluing equations are not unique we may still define $L^r$ as the minimum value reached by $L$ for the given boundary conditions. 
In our experience with systems in $1$-dimensional spacetimes, we saw that 
for a particle moving in euclidian space under a potential the 
gluing equation has a unique solution, while for the rigid body in discrete time, the solution to the gluing equation is not unique. 
When solutions of the gluing equation are not unique, one should verify that 
the definition of $L^r$ is physically appropriate. For example, the sign in the 
original action for 
Veselov's model of rigid body motion had to be modified for this definition to work. 
We will also find gluing equations with 
double solutions and explicitly define the reduced model in 
the lattice gauge theory example presented in Subsection \ref{lgt}. 
Clearly this issue is delicate in general situations, but the simplicity of the gluing equations should make a meaningful analysis possible in most cases. The issues to investigate are if in the case of multiple solutions there are some that can be categorized as physical and some as discretization artifacts, and if the reduced lagrangian $L^r$ is well defined. 

Once the action is written we can proceed to investigate the geometrical structures that follow from the variational problem. The reduced system takes place in a discretization in which for each dual cell, the ``boundary degrees of freedom'' are not shared by only two cells. This complicates the organization of degrees of freedom to define the structural forms; in order to achieve this task one would have to follow conventions of the type given by Marsden et al. \cite{MarsdenEtAl}.

\section{Local covariant hamiltonian picture}
% - The canonical framework
\label{Canonical}

The multisymplectic framework for field theory 
allows for a useful separation into kinematical structure and dynamics. 
This separation does not happen in the lagrangian picture on the jet bundle $J^1Y$ that we have been using; the separation occurs in the hamiltonian picture on the dual jet bundle 
$J^1Y^\ast$. In that space there are canonical forms $\Theta$, $\Omega = -d\Theta$. 
We will not attempt to produce a new map that let us pull back those canonical forms from the continuum to our discrete setting. 
If we had constructed the discrete theory through regularization in the manner described in Section (\ref{Reg-Coar-Cont}), 
the data of our discrete spacetime model would be converted into histories and variations in the continuum.
Once in the continuum, in $J^1Y$, the continuum lagrangian 
induces a covariant Legendre transformation \cite{MarsdenEtAl} which brings the structure to the lagrangian setting. 
Finally, we can pull back the canonical structure all the way to the discrete lagrangian framework using the regularization map. 

Here we will develop the most basic elements of a covariant canonical framework for field theory on discrete spacetime 
starting from the geometry of the discrete field theory that we have already built.

The jet bundle $J^1Y$ is the space of equivalence classes of local sections $\phi:Y \to M$, that agree up to first order; $[\phi(x)] \in J^1Y$ is composed of local sections evaluated on $x\in M$ with the value $\phi(x) \in Y$ and that have the same partial derivatives. 
$J^1Y^\ast$ is a bundle over $Y$ whose fiber over $\phi(x) \in Y$ is composed of affine maps from the fiber over $\phi(x) \in Y$ to the space of $n$-forms on $x\in M$. 
We proposed a discrete analog of $J^1Y$; in the case of scalar fields, discrete local sections that agree up to first order at $\nu$ are written as 
\[
\tilde{\phi}(\nu) = 
(\nu, \phi_\nu \in \cal{F}, \{ \phi_\tau \in \cal{F} \}_{\tau \subset \partial \nu} ) . 
\]
Once $\nu$ and $\phi_\nu$ are fixed, the specification of the equivalence class of agreement up to first order is done with $\{ \phi_\tau \}_{\tau \subset \partial \nu}$. 
Given a pair 
$(\phi_\nu, \phi_\tau)$, we 
evaluate the change of the local section as one moves 
from $C\nu$ in 
the direction of $C\tau$ by means of a translation vector 
$V(\phi_\nu , \phi_\tau) [\phi_\nu] = \phi_\tau$, and 
we assume that 
in the fiber the vector space of translations acts transitively,%
\footnote{
We could also treat the case were the admitted transitive translations are right group translations, but we will not do it here. The reader interested in nonlinear sigma models can infer the formalism from the case of the scalar field and the case of gauge theories presented at the end of this section.
} 
which implies   
$V(\phi_\nu , \phi_\tau)= \phi_\tau - \phi_\nu$. 
However, in order to mirror the gluing properties of our lagrangian formalism, we will work with the collection of estimates of the change in the local section from each $C\tau$ as one moves to $C\nu$. The relations are $V(\phi_\tau, \phi_\nu) [\phi_\tau] = \phi_\nu$; the important aspect is that the value of $\phi_\tau$ is shared by neighboring the atoms intersecting at $\tau$. 
We will prescribe maps from this space of classes to the space of cochains $\Lambda \nu^\ast$ (with $\Lambda \in \R$) that act on $\nu$. 
We parametrize affine maps from the space of classes of local sections to the space of cochains by 
\[
\bar{\psi}(\nu) = 
( \nu, \psi_\nu, p_\nu \in \R, 
\{ (\psi_\tau , p_\tau) \in T\ast \cal{F}\}_{\tau \subset \partial \nu} ) . 
\]
We will also write 
$\bar{\psi}(\nu_\tau) = ( \nu, \psi_\nu, p_\nu , (\psi_\tau , p_\tau))$. 
If $\psi_\tau = \phi_\tau$, we can calculate 
\[
\bar{\psi}(\nu_\tau) [\tilde{\phi}(\nu)] = 
\left( \frac{p_\nu}{n+1} + p_\tau d\psi_\tau [V(\phi_\tau, \phi_\nu)] \right) 
\nu^\ast . 
\]
Notice that $\bar{\psi}$ contains all the information about a section $\psi$, which may be written in the first order format as $\tilde{\psi}$. In this way,  $\bar{\psi}$ can ``act on itself'' giving the $n$-cochain $\bar{\psi} [\tilde{\psi}]$. 
This structure will be very useful; 
for example, it gives us a measure in $U$. 
In our discrete spacetime setting $F: U_{\tiny\mbox{disc}} \to \R$ is thought of as the decimation of a function in the continuum $F_{\tiny\mbox{cont}}:M \to \R$. 
The measure $\bar{\phi} [\tilde{\phi}]$ gives 
\[
\sum_U F 
\bar{\phi} [\tilde{\phi}] = 
\sum_{U_{\tiny\mbox{disc}}}
F_\nu ( p_\nu + \sum_{\tau \subset \partial \nu} 
p_\tau d\phi_\tau [V(\phi_\tau, \phi_\nu)] ).
\]

Our notation for variations is 
$\bar{v}(\nu) = (v_{\psi_\nu}, v_{p_\nu}, \{ (v_{\psi_\tau} , v_{p_\tau}) \}_{\tau \subset \partial \nu} )$. 

Throughout this article, our notation has hinted at the idea of $\tilde{\phi}(\nu)$ 
and $\tilde{\phi}(\tau)$ representing an $n$-chain and an $n-1$-chain in the discrete jet bundle, the images of $\nu$ and $\tau$ under $\tilde{\phi}$. In the same way, we may think of 
$\bar{\phi}(\nu)$ and $\bar{\phi}(\tau)$
as an $n$-chain and an $n-1$-chain in the discrete dual jet bundle. 
From the form of the map given above, we can infer the formulas for the 
canonical forms/cochains 
\[
\hat{\Theta} (\bar{\phi}(\nu)) = p_\nu + \sum_{\tau \subset \partial \nu} 
p_\tau d\phi_\tau [V(\phi_\tau, \phi_\nu)]  , 
\]
\[
\Theta (\bar{v}, \bar{\phi}(\tau_\nu)) 
= 
p_\tau d\phi_\tau [v_{\phi_\tau}] , 
\]
where the subindex $\nu$ in the $n-1$-chain $\tau_\nu$ is necessary because $\tau$ is shared by two $n$ simplices, and we need to specify which one of them is used to evaluate. 
Clearly, we can also define $\hat{\Omega} = -d\hat{\Theta}$ and 
$\Omega = -d \Theta$. 

The covariant Legendre transformation relating the lagrangian and hamiltonian pictures is $f_L: 
U_\Delta^n \times{\cal F}_\nu \times (\times_{\tau' \subset \partial \nu} {\cal F}_{\tau'})
\to 
U_\Delta^n \times{\cal F}_\nu \times 
(\times_{\tau' \subset \partial \nu} T^\ast {\cal F}_{\tau'})$ 
prescribed by 
\[
f_L(\tilde{\phi}(\nu)) = 
( \nu, \phi_\nu, p_\nu , 
\{ (\phi_\tau , p_\tau ) \}_{\tau \subset \partial \nu} )
\]
with $p_\nu = L(\tilde{\phi}(\nu))
- \sum_{\tau \subset \partial \nu} 
\frac{\partial L}{\partial \phi_\tau}(\tilde{\phi}(\nu))
d\phi_\tau [V(\phi_\tau, \phi_\nu)]$ and $p_\tau = \frac{\partial L}{\partial \phi_\tau}(\tilde{\phi}(\nu))$. 

The relation between the lagrangian and the canonical structures is 
\[
f_L^\ast \hat{\Theta} = L  , \quad 
f_L^\ast  \Theta = \Theta_L   , \quad 
f_L^\ast \hat{\Omega} = \hat{\Omega}_L  , \quad 
f_L^\ast  \Omega = \Omega_L  . 
\]

Notice that the dimension of the discrete version of the dual jet bundle is bigger than that of the discrete version of the jet bundle, as it happens in the continuum. Also, in the discrete dual jet bundle there are canonical structure forms/cochains $\hat{\Theta}, \Theta$ and from the point of view of this hamiltonian formalism all the information from the lagrangian is collected in the image of the Legendre transform that is a proper submanifold of the discrete dual jet bundle. Thus, in this covariant hamiltonian framework all the information is contained in {\em constraints}, relations that define the locus of the dual jet bundle that is relevant for the system (the image of the Legendre transform). 

The geometry of our version of dual jet bundle on discrete spacetime is 
therefore 
related to the variational problem posed earlier within the lagrangian picture of field theory on discrete spacetime. 

Solutions to our lagrangian field equations, when mapped to the dual bundle using $f_L$, have the interpretation of being constructed using the lagrangian as generating function. Let us explain. 
We start the study with a single corner cell $c$; physical motions over $c$ in the hamiltonian picture $\bar{\phi}|_c = f_L(\tilde{\phi})|c$ 
are constructed solving implicit equations using the lagrangian $L^c$ as generating function 
(and the definitions of momenta in terms of the partial derivatives of the lagrangian given above). In a spacetime atom $\nu$ several corners meet, and there are gluing conditions for the solutions over them. 
Consider a local hamiltonian history of the form $\bar{\phi}(\nu) = f_L(\tilde{\phi}(\nu))$ and variations of the form $\bar{v}(\nu) = f_{L\, \ast}\tilde{v}(\nu)$. 
The corner pasting conditions for the history in the interior of the atom are 
$0 = \sum_{\nu - \partial \nu} \tilde{\phi}^\ast ( \tilde{v} \lrcorner \hat{\Omega}_L) =  
\sum_{\nu - \partial \nu} \bar{\phi}^\ast ( \bar{v} \lrcorner \hat{\Omega})$ for every 
$\tilde{v}$. 
The same procedure can be used to glue solutions over corner cells from neighboring atoms bearing in mind momentum matching conditions on the face shared by the two neighboring atoms. 
In this sense, the discrete lagrangian $L$ is a local covariant generating function; in order to construct solutions using it we need the definitions of momenta given above and local gluing conditions of the type described above. 
All the local gluing conditions for $\bar{\phi}$ are contained in the equation 
$0 = \sum_{U - \partial U}  \bar{\phi}^\ast \hat{\Omega}$, which is read as the vanishing of a differential $1$-form acting on hamiltonian variations of the type $\bar{v}$ of the type $f_{L\, \ast}\tilde{v}$. 

The existence and uniqueness of solutions to this set of coupled equations is an issue of fundamental importance. 
Detailed studies of related existence and uniqueness problems 
have been carried on, see for example  \cite{GambiniETAL, DittrichETAL}. 
We could also pursue a hamiltonian picture in terms of pure boundary data, the analog of what is presented in Subsection \ref{BdaryDataF}. 
In this picture of the boundary framework solutions over each atom $\nu$ are constructed implicitly using the boundary lagrangian as generating function without any restriction; then local solutions are amalgamated by gluing conditions on shared faces. 
We will not describe this formalism further. 

In the space of solutions of the 
lagrangian picture the multisymplectic formula holds, 
$0 = \sum_{\partial U} \tilde{\phi}^\ast \Omega_L$; thus, in the space of physical motions of the hamiltonian picture we have a similar structural conservation law, 
$0 = \sum_{\partial U} \bar{\phi}^\ast \Omega$. 
Similarly, Noether's theorem is stated in the covariant canonical framework 
in terms of the momentum map provided by the canonical form $\Theta$. 
When the lagrangian has symmetries the image of the Legendre transform is even smaller due to relations among the momenta. This issue will play a predominant role when we deal with gauge fields.

Now we will write down the basic elements the canonical formalism 
for the case of {\em modified BF theories}, and we will not treat the case of {\em gauge fields} separately because gauge fields are part of the degrees of freedom present in modified BF theories; thus, the canonical structure in models for gauge fields is somehow contained in that of models for modified BF theories. 
In this paragraph, 
our treatment will specialize in $G=SO(N)$, but the case of $SU(N)$ proceeds in entire analogy. 
We remark that our treatment of different types of fields may differ due to the way in which they appear in the lagrangian. For example, the lagrangians that we consider for modified BF theories do not include partial derivatives of the $e$ field. 

Histories in the first order format (classes of local sections up to discrete first order agreement) 
were written as 
$\tilde{\ae}(\nu) =(\nu, \{ h_l \}_\nu , \{ k_r \}_\nu , \{ e_s \}_\nu , \varphi_\nu)$. In this section we will change the order of the coordinates by writing the interior degrees of freedom first and at the end the degrees of freedom associated with the boundary faces, 
$\tilde{\ae}(\nu) =(\nu; \varphi_\nu , \{ e_s \}_\nu , \{ h_l \}_\nu ; \{ k_r \}_\nu )$. 

The only field whose derivatives appear in the lagrangians of interest is the connection. 
Gluing is more powerful if it is done at the co-dimension two simplices $\sigma$ because many wedges join there. Thus, from each $\sigma \subset \nu$ we consider classes of discrete sections (connections) that agree at up to first order around $C\sigma$ and affine maps from those classes to the space of $n$-cochains. 
We are interested in describing how the section (the connection) changes from the boundary of an atom $\nu$ as we move towards the interior. 
In our discrete model there is an object that tells us how parallel transport changes as we move a path from the boundary towards the interior of $\nu$ living the end points fixed. 
We define $\hat{g}_{\partial s}$ as the parallel transport along $\partial s$ with base point $C\sigma$, and we consider an orientation on $s$ such that is compatible with the orientation of $r$. 
Recall that 
$k_{r2}^{-1} k_{r1}$ records the parallel transport from $C\tau_1$ to $C\sigma$ to 
$C\tau_2$ along $r_2^{-1}\circ r_1$. 
On the other hand, 
the expression  
$k_{r2}^{-1} \hat{g}_{\partial \bar{s}} k_{r1}= h_{l2} h_{l1}^{-1}$ gives the parallel transport from $C\tau_1$ to $C\tau_2$ through the route in $-\partial s= \partial \bar{s}$ that goes trough the interior of $\nu$. 
This transformation may be seen as $\hat{g}_{\partial \bar{s}}$ acting on $k_{r1}$ by left translation.
Analogous infinitesimal displacements of $k_{r1}$ are written as 
$\eta k_{r1} \in T_{k_{r1}} G$. We will use an orthonormal basis in the Lie algebra to shorten some expressions; 
we will write $\eta = \eta^i f_i$. 
Thus, points in the discrete dual bundle are written as 
\[
\nu \overset{\bar{\AE}}{\longmapsto} 
(\nu; \varphi_\nu , \{ e_s \}_\nu , \{ h_l \}_\nu ; p_\nu \in \R ; 
\{ ( k_r , p_r) \in T^\ast G \}_\nu ) . 
\]
The covector $( k_r , p_r)$ acts naturally on $\eta k_r \in T_{k_r} G$, and it can also act on the ``large'' 
left translations of $k_r$ introduced above. Here is how: 
\[
p_r [\eta k_r ] = -\Tr(k_r^T 
p_r
\eta k_r )= p_{r\, i} \, \eta^i , \quad
p_r [\hat{g}_{\partial \bar{s}} k_r] 
= -\Tr( k_r^T 
p_r 
\hat{g}_{\partial \bar{s}} k_r)
= - p_{r\, i} \, \hat{\theta}_s^i ,
\]
where we have defined  
$p_r=p_{r\, i} \, f^i$, 
$\hat{\theta}_s^i= -\Tr( f^i \hat{g}_{\partial s})$. 

If $\bar{\AE}$ and $\tilde{\ae}$ give the same $k_r$, we can calculate
\[
\bar{\AE}(\nu) [\tilde{\ae}(\nu)] = 
\left( \frac{p_\nu}{n(n+1)} - p_{r\, i} \, \hat{\theta}_s^i 
\right) 
\nu^\ast . 
\]
Following the procedure used above for the case of the scalar field we find 
\[
\hat{\Theta} (\bar{\ae}(\nu)) = p_\nu - \sum_{s \subset \nu} 
(p_{r_s} - p_{r'_s})_i
\hat{\theta}_s^i  , \quad 
\Theta (\bar{v}_\eta, \bar{\phi}(\tau_\nu)) 
= \sum_{r \subset \tau} 
p_{r\, i} \, \eta^i   
\]
where $\bar{r'_s} \circ r_s$ is the part of $\partial s$ contained in $\partial \nu$. 

The covariant Legendre transformation is 
\[
f_L(\tilde{\ae}(\nu)) = 
(\nu; \varphi_\nu , \{ e_s \}_\nu , \{ h_l \}_\nu ; p_\nu \in \R ; 
\{ ( k_r , p_r) \in T^\ast G \}_\nu )
\]
with 
$p_{r\, i}  = \frac{\partial L}{\partial k_r}(\tilde{\ae}(\nu)) dk_r [f_i k_r]$ and 
$p_\nu = L(\tilde{\ae}(\nu)) 
+ \sum_{s \subset \nu} 
(p_{r_s} - p_{r'_s})_i \hat{\theta}_s^i $. 
Hamiltonian and lagrangian structures are related by 
\[
f_L^\ast \hat{\Theta} = L  , \quad 
f_L^\ast  \Theta = \Theta_L   , \quad 
f_L^\ast \hat{\Omega} = \hat{\Omega}_L  , \quad 
f_L^\ast  \Omega = \Omega_L  . 
\]
A similar relation holds for the gauge theories presented in this article. 

\vskip0.2cm
\noindent
{\em Remarks} \\
First, the discrete version of the dual jet bundle that we just presented is composed of a collection of cotangent bundles. The reader may have noticed that, in the case of the scalar field, there is no copy of $T^\ast {\cal F}$ associated with the degree of freedom $\phi_\nu$. Our geometric construction is cleaner without it. 
On the other hand, if within the hamiltonian picture we wanted to derive the reduced formalism where the boundary degrees of freedom are eliminated, we would need the omitted copy of the cotangent bundle. The origin of this issue lies in our lagrangian formalism: our discrete version of Cartan's form $\Theta_L = f_L^\ast \Theta$ acted on a pair composed of a vector and a $n-1$-chain of the type $\tilde{\phi}_\tau$. 
Allowing discrete $n-1$-surfaces that cut through the interior of spacetime atoms is also possible; in this extended framework, $\Theta_L$ acting on a $n-1$-chain touching $C\nu$ would involve 
$p_{\phi_\nu}=\frac{\partial L}{\partial \phi_\nu}(\tilde{\phi}(\nu))$. \\
Second, we notice that the construction of solutions by means of gluing covariant generating functionals has the  simple gluing property; that is, the gluing conditions are simply matching of momenta calculated from the atoms that intersect along a face as if the theory were the free theory. 
The third observation is that if the lagrangian of our model is of the form of a free theory plus an interaction term, with the interaction term secluded to interior degrees of freedom, then 
the covariant Legendre transformation does not involve partial derivatives of the interaction term. From the point of view of the canonical formalism, this is the reason that we have simple gluing in the lagrangian picture. \\
Third, we saw that in the lagrangian picture symmetries implied redundancies among the equations of motion and that local gauge symmetries actually implied relations that constrained atomic boundary data. In the hamiltonian picture all this has its counterpart, and it is is aided with the geometrical clarity that comes with the canonical structure. 
In this context gauge symmetries are local (inside atoms) group actions that 
preserve the constraint surface and that have ho physical meaning. The canonical form $\Theta$ gives us conserved quantities that have the interpretation of redundancies among the relations defining the constraint surface. In other words, local gauge symmetries are responsible for part of the relations defining the constraint surface. \\
Fourth, 
this covariant canonical formalism on discrete spacetime 
should be explored further; here we focus on 
laying down the basic elements and showing its relation with the lagrangian formalism that we developed in this paper.

\section{Examples}
\label{FTexamples}
% 4) Examples (  ,  ,  ,Reisenberger's model)
%
\subsection{Nonlinear waves}
\label{nlwaves}
Apart from the physical interest of this system, we present it as an example in this article in order to compare the resulting model with the model developed by Marsden, Patrick and Shkoller in \cite{MarsdenEtAl}.
The main difference is due to the extra structure present in our discretization of spacetime. 
In order to get a closer comparison of the two models, we develop a 
reduced model solving the gluing field equations, and we also comment on a model based on the boundary data of our spacetime atoms. 

We consider nonlinear waves in two-dimensional Minkowski space described by the action $S(\phi)= \int_U \left\{  
\frac{1}{2} \left[ \frac{\partial \phi}{\partial x^0}^2 - \frac{\partial \phi}{\partial x^1}^2 \right] 
+ N(\phi) \right\} dx^1 \wedge dx^0$. 
Our model is based on a regular cartesian cellular decomposition. 
Given an atom $\nu$ its faces will be denoted by pairs
$(\tau=0+, \tau=0-, \tau=1+, \tau=1-)$; for example, the face $\tau=0+$ is the one shared with the neighboring atom in the positive $x^0$ direction. 
The discrete action is $S(\phi)= \sum_U L(\tilde{\phi}(\nu))$, with 
the discrete lagrangian being a sum of corner lagrangians 
$L=\sum_c L^c = L^{++} + L^{+-} + L^{-+} + L^{--}$. The lagrangian for the corner 
corresponding to increasing $x^0$ and $x^1$ directions is 
\[
L^{++}(\tilde{\phi}(\nu)) = \left\{ 
\frac{1}{2} \left[ \left(\frac{\phi_{0+} -\phi_\nu}{h}\right)^2 - 
\left(\frac{\phi_{1+} -\phi_\nu}{k}\right)^2 \right] 
+ N(\phi_\nu) \right\}hk , 
\]
where $h$ is the modulus of the interval between $C\nu$ and $C_{0+}$ (and $C_{0-}$) and $k$ is the modulus of the interval between $C\nu$ and $C_{1+}$ (and $C_{1-}$). 
The other corner lagrangians are constructed similarly. 
One must be careful with the sign in the difference operators that regularize derivatives; for example, in $L^{+-}$ 
the difference operator regularizing a derivative in the $x^1$ direction is not the one used in $L^{++}$, but $\phi_\nu - \phi_{1-}$. 

There are two types of field equations: (i) Equations of the first type, $0 = \frac{\partial L}{\partial \phi_\nu}(\tilde{\phi}(\nu))$, ensure that $\phi$ is a solution inside an atom $\nu$. In this case they are 
\[
 0=2hk \left(
- \frac{\phi_{0+} -2\phi_\nu + \phi_{0-}}{h^2}
+ \frac{\phi_{1+} -2\phi_\nu + \phi_{1-}}{k^2}
+ 2 N'(\phi_\nu)
\right) . 
\]
(ii) Equations of the second type, $0 =\frac{\partial L}{\partial \phi_\tau}(\tilde{\phi}(\nu)) 
+ \frac{\partial L}{\partial \phi_\tau}(\tilde{\phi}(\nu'))$, 
are gluing conditions for solutions over two neighboring atoms $\nu, \nu' \subset U$. 
If $\nu'$ is a future time translation of $\nu$, the 
explicit form of the gluing condition is 
\[
 0=2k \left(
\frac{\phi_{0+}(\nu) - \phi_\nu}{h}
- \frac{\phi_{\nu'} - \phi_{0-}(\nu')}{h}
\right) .  
\]
On the other hand, if $\nu'$ is a positive translation in direction $x^1$ of $\nu$, the 
explicit form of the gluing condition is 
\[
 0=-2h \left(
\frac{\phi_{1+}(\nu) - \phi_\nu}{k}
- \frac{\phi_{\nu'} - \phi_{1-}(\nu')}{k}
\right) .  
\]
Notice that these equations (in both cases) are solved for $\phi_\tau$ setting 
$\phi_\tau = \frac{\phi_\nu + \phi_{\nu'}}{2}$ for $\tau = \nu \cap \nu'$. 
Alternatively, if the unknown is $\phi_{\nu'}$ the solution is 
$\phi_{\nu'}= 2 \phi_\tau - \phi_\nu$. 

In this case the structural forms 
$\Theta_L(\cdot , \tilde{\phi}(\tau_\nu)), 
\Omega_L(\cdot , \cdot , \tilde{\phi}(\tau_\nu))
$
are 
\[{\scriptstyle
\Theta_L(\cdot , \tilde{\phi}(1+_\nu)) = 
-2h \frac{\phi_{1+}(\nu) - \phi_\nu}{k} d\phi_{1+}
\quad , \quad 
\Theta_L(\cdot , \tilde{\phi}(0+_\nu)) = 
2k \frac{\phi_{0+}(\nu) - \phi_\nu}{h} d\phi_{0+} ,}
\]
\[{\scriptstyle
\Theta_L(\cdot , \tilde{\phi}(1-_\nu)) = 
2h \frac{\phi_\nu - \phi_{1-}(\nu)}{k} d\phi_{1-}
\quad , \quad 
\Theta_L(\cdot , \tilde{\phi}(0-_\nu)) = 
-2k \frac{\phi_\nu - \phi_{0-}(\nu)}{h} d\phi_{0-} ,}
\]
\[{\scriptstyle
\Omega_L(\cdot , \cdot , \tilde{\phi}(1+_\nu)) = 
-\frac{2h}{k} d\phi_\nu \wedge d\phi_{1+}
\quad , \quad 
\Omega_L(\cdot , \cdot , \tilde{\phi}(0+_\nu)) = 
\frac{2k}{h} d\phi_\nu \wedge d\phi_{0+} ,}
\]
\[{\scriptstyle
\Omega_L(\cdot , \cdot , \tilde{\phi}(1-_\nu)) = 
-\frac{2h}{k} d\phi_\nu \wedge d\phi_{1-}
\quad , \quad 
\Omega_L(\cdot , \cdot , \tilde{\phi}(0-_\nu)) = 
\frac{2k}{h} d\phi_\nu \wedge d\phi_{0-} .}
\]
This geometric structure leads to a very simple multisymplectic formula even when the nonlinearity makes finding solutions a great or impossible challenge. 

Now we will use the field equations in two ways. Firstly, we will comment on solving the time evolution problem given initial data. 
Secondly, we will use ``bulk variables'' $\{ \phi_\nu\}_{\nu \in U}$ to determine histories; the rest of the variables will be generated solving the gluing equations. 
A reduced model, with fewer variables and fewer equations, will be studied and compared with other approaches. 

\noindent
{\em On the evolution problem}\\ 
Consider $\phi'$ a solution of the equations on a region $U^0 \subset U$ that is identical to $U$ to the past of an isochronous hypersurface $\Sigma^0$. In our example $\Sigma^0$ is one-dimensional, and we assume that it is 
decomposed into 
faces $\tau$ where a past atom $\nu$ meets a future atom $\nu'$. Our formula, $\phi_{\nu'}= 2 \phi_\tau - \phi_\nu$, applied to all $\tau \subset \Sigma^0$, generates new data to the future of $\Sigma^0$. This data 
$\{ \phi_{\nu'}\}_{\tiny\mbox{New}}$ can be entered in the gluing equations 
$\phi_{\tau'} = \frac{\phi_{\nu'} + \phi_{\nu''}}{2}$ to complete a solution in a bigger domain; note that the boundary conditions enter only in the equations associated to history variables on atoms that touch $\partial U$. 
The bigger domain 
$\hat{U^0}\supset U^0$ is identical to $U$ to the past of an isochronous hypersurface $\hat{\Sigma^0}$. 
Over each face $\tau$ of 
$\Sigma^0 \cap U$ 
there is an atom $\nu'$ and we have extended the solution {\em halfway} on each $\nu'$; we still need to find $\phi_{0+}(\nu')$ in order to evolve one complete step towards the future. For each $\nu'$, the equations of type (i) give us $\phi_{0+}(\nu')= 2\phi_{\nu'} - \phi_{0-}(\nu') + 
h^2 (\frac{\phi_{1+}(\nu') -2\phi_\nu + \phi_{1-}(\nu')}{k^2} + 2N'(\nu'))$. 
With this new data we have extended the solution $\phi$ one step to the future. Our extended solution is now valid in the region $U^1 \subset U$ that is identical to $U$ to the past of an isochronous hypersurface $\Sigma^1$. We could iterate this process to find solutions from initial data. Stability issues will not be discussed in this article. 

If we are interested in the evolution problem from Cauchy data on a non compact Cauchy surface, the formalism of this article needs to be appropriately complemented.

\noindent 
{\em The reduced model} \\
We are given the value of ``atomic bulk variables'' $\phi^r=\{ \phi_\nu\}_{\nu \subset U}$. This data can be used to determine a complete history $\phi(\phi^r)$ 
by means of the formula $\phi_\tau = \frac{\phi_\nu + \phi_{\nu'}}{2}$ for $\tau=\nu \cap \nu'$. The constructed history, $\phi(\phi^r)$, solves the gluing equations automatically. Equations of type (i) remain unsolved and can be written in terms of bulk variables leading to 
\[
0 = hk \left(
- \frac{\phi_{\nu+e_0} -2\phi_\nu + \phi_{\nu-e_0}}{h^2}
+ \frac{\phi_{\nu+e_1} -2\phi_\nu + \phi_{\nu-e_1}}{k^2}
+ 4 N'(\phi_\nu) \right) . 
\]
We can also evaluate the action $S$ on the solutions constructed from the ``atomic bulk variables''. 
Solutions generated in this way do not cover the whole region $U$ because the variables at $\partial U$ cannot be calculated. We then consider a slightly contracted domain $U_r \subset U$ that has points of the type $C\nu$ in its boundary. 
The reduced action is not difficult to calculate; the only new element to consider is that there are atoms whose interior intersects $\partial U_r$. Thus, in the sum 
$S_r(\phi^r)= \sum_{Ur} L(\widetilde{\phi(\phi^r)}(\nu))$ the terms touching the boundary use lagrangians in which only some of the corner lagrangians participate. 
The corner lagrangian $L^{++}$ in terms of the reduced variables is 
\[
L^{++}(\widetilde{\phi(\phi^r)}(\nu)) = \left\{ 
\frac{1}{2} \left[ \left(\frac{\phi_{\nu+e_0} -\phi_\nu}{2h}\right)^2 - 
\left(\frac{\phi_{\nu+e_1} -\phi_\nu}{2k}\right)^2 \right] 
+ N(\phi_\nu) \right\}hk ,
\]
where $\nu+e_0$ denotes the atom neighboring to $\nu$ in the $+e_0$ direction; in this way we can name all the neighbors of $\nu$ as $\nu+e_1, \nu+e_0, -\nu+e_1, -\nu+e_0$. The other corner lagrangians have an analogous form. 

We can work with the reduced histories $\phi^r$ using a notation natural for a cartesian grid. Points $C\nu$ will be denoted by a pair of integers $(i, j)$, and the action will be reorganized as a sum over plaquettes determined by their corners 
$\Box=((i, j), (i, j+1), (i+1, j+1), (i, j+1))$. The plaquette lagrangian is also a sum over corner lagrangians $L^\Box=\sum_c L^c$, but a different sum from the one used in the main part of this article. A portion of a reduced history over a plaquette will be written as 
$\tilde{\phi^r}(\Box)= (\phi_{i, j}, \phi_{i, j+1}, \phi_{i+1, j+1}, \phi_{i, j+1})$. With this notation the reduced action is 
$S_r(\phi^r)= \sum_{Ur} L^\Box(\tilde{\phi^r}(\Box))$. The explicit form of the plaquette lagrangian in this example $L^\Box(\tilde{\phi^r}(\Box))$ is 
\begin{eqnarray*}
\Bigg\{ \Bigg.
\frac{1}{2} \!\!\!\!
&& \!\!\!\! \Bigg[ 
\frac{
\big(\frac{\phi_{i+1,j}-\phi_{i,j}}{2h}\big)^2 
+ 
\big(\frac{\phi_{i+1,j+1}-\phi_{i,j+1}}{2h}\big)^2 
}{2} 
- 
\frac{
\big(\frac{\phi_{i, j+1} -\phi_{i,j}}{2k}\big)^2 
+ 
\big(\frac{\phi_{i+1, j+1} -\phi_{i+1,j}}{2k}\big)^2
}{2} 
\Bigg] \\
&& + 
\Bigg. \frac{1}{4}\Big(N_{i, j}+N_{i, j+1}+N_{i+1, j+1}+N_{i, j+1}\Big) 
\Bigg\}4hk  ,
\end{eqnarray*}
where we wrote 
$N(\phi_{i, j})$ as $N_{i, j}$. 

We have written explicitly the action and the field equations of the reduced model. If the reader compares them with the model of Marsden, Patrick and Shkoller  \cite{MarsdenEtAl}, he or she will find that they differ. The comparison would be complete if the geometric structures were contrasted. In contrast to the structure of our non-reduced formalism, 
the resulting structure explicitly depends on $N$.

\noindent 
{\em A model based on atomic boundary data} \\
Reducing by solving the interior field equations may be impossible, or may produce a cumbersome model. Instead of reducing, it is natural to start with a new regularization that uses only atomic boundary data. A local first order history is written as 
$\tilde{\phi}^b(\nu) = (\nu, \phi_{0+}, \phi_{0-}, \phi_{1+}, \phi_{1-})$, and a reasonable regularization yields 
%the discrete lagrangian 
$L^b(\tilde{\phi}^b(\nu))$ equal to 
%whose explicit evaluation is 
\[
\left\{ 
\left(\frac{\phi_{0+} -\phi_{0-}}{h}\right)^2 - 
\left(\frac{\phi_{1+} -\phi_{1-}}{k}\right)^2 + 
N\left( \frac{\phi_{0+} + \phi_{0-} + \phi_{1+} +\phi_{1-}}{4} \right)
\right\} 4 hk . 
\]
The functional form of this discrete lagrangian looks like the model proposed by 
Marsden, Patrick and Shkoller  \cite{MarsdenEtAl}, but the connectivity of the lattice is different. The argument of this discrete lagrangian consists of atomic boundary data. A site is a representative of a face shared by two atoms; in comparison, in a cartesian two dimensional lattice each site has four neighbors. 
The resulting structure has the disadvantage of explicitly depending on $N$.

\subsection{Lattice gauge theory without fermions}
\label{lgt}
We apply the formalism of 
Subsection \ref{gauge}
to a regularization of euclidean pure Yang-Mills theory on a cubical cellular decomposition of a domain $U$ in spacetime. 
Recently Halvorsen, Sørensen and Christiansen developed a version of 
Noether's theorem for spacetime simplicial gauge theory \cite{NoetherLGT}. 
Their formalism is related to part of our formalism, but they use a different discretization. We present this example on a cubical lattice to avoid much of the needed regularization work due to the shape of the elements of the discretization. 
If the reader is interested in a version of this example on a triangulation, the first step would be to read their paper and also recent work on lattice regularization for general field theories \cite{LatticeRegularization}. 
The version of this model for lorentzian signature does not look very different at the level of the general formalism. Of course, the radical difference 
lies in the solutions to the equations. 
We present the euclidian version hereto show that our formalism is not restricted to hyperbolic PDEs. 

We use a cubical cellular decomposition.  
The action for this regularization of euclidean pure Yang-Mills theory is 
\[
S_{\rm Euc}(A)= \beta
\sum_{\nu \subset U} \sum_{s<\nu} 
[1 - \frac{1}{N} \Rp \Tr(g_{\partial s})] , 
\]
where $\beta$ is a free parameter, and 
we have assumed that $G=SU(N)$. The definition of the action needs an orientation in each wedge $s$ (the shaded area in Figure \ref{fig}.b) to write 
$g_{\partial s}= h_{l2}^{-1} k_{r2}^{-1} k_{r1} h_{l1}$, but the choice is irrelevant for the calculation. This action is the analog of Wilson's action \cite{Wilson} 
in our more structured discretization. 

In order to study the variation of the action we parametrize variations as 
$\tilde{v}_\xi(\nu)= (\{ h_l\xi_l \in T_{h_l}G \}_\nu , 
\{ \xi_rk_r \in T_{k_r} \}_\nu)$, where we have denoted elements of the Lie algebra by $\xi$, and we treat them as antihermitian matrices. Furthermore, we choose a basis in the Lie algebra to write $\xi= \xi^i f_i$ where the coefficients are real numbers. 
The variation of the action involves derivatives of two types: The first type is  
$\frac{\partial}{\partial h_{l1}}
(\frac{-\beta}{N}\Rp \Tr(g_{\partial s}))dh_{l1}[h_{l1}\xi_{l1}]= \frac{-\beta}{N}\Rp \Tr(g_{\partial s}\xi_{l1})= \frac{\beta}{N}\xi_{l1}^i \theta_{s, i}$, where we have defined 
$\theta_{s, i} = - \Rp \Tr(f_ig_{\partial s})$. The second type 
is $\frac{\partial}{\partial k_{r1}}
(\frac{-\beta}{N}\Rp \Tr(g_{\partial s}))dk_{r1}[\xi_{r1}k_{r1}]
= \frac{\beta}{N}\xi_{r1}^i \hat{\theta}_{s, i}$, where we have defined 
$\hat{\theta}_{s, i} = -\Rp \Tr(f_i k_{r1} h_{l1} h_{l2}^{-1} k_{r2}^{-1})$. 
The resulting variation of the action may be written as 
$dS(A) [v] =
- \sum_{U -\partial U}
\tilde{A}^\ast(\tilde{v} \lrcorner \hat{\Omega}_L) 
+ 
\sum_{\partial U} 
\tilde{A}^\ast(\tilde{v} \lrcorner \Theta_L)$, with 
\[
\Theta_L(\tilde{v}_\xi , \tilde{A}(\tau_\nu)) = 
\tilde{v}_\xi L(\tilde{A}(\nu))|_\tau = 
\frac{\beta}{N}\sum_{r \subset \tau} \xi_r^i \hat{\theta}_{s, i} , 
\]
\[
\hat{\Omega}_L(\tilde{v}_\xi , \tilde{A}(\nu)) = 
-\tilde{v}_\xi L(\tilde{A}(\nu)) = 
-\frac{\beta}{N}\sum_{l \subset \nu} \xi_l^i \sum_{s \supset l} \theta_{s, i}
-\frac{\beta}{N}\sum_{r \subset \partial\nu} \xi_r^i \hat{\theta}_{s(r), i} .
\]
The field equations interior to an atom $\nu$ are 
\[
\sum_{s \supset l} \theta_{s, i}=0 \mbox{ for every } l \subset \nu , 
\]
where the orientation of each $s$ is such that the orientation of $\partial s$ agrees with the orientation of $l$. 
The gluing field equations for $\tau = \nu \cap \nu'$ with $\nu, \nu' \subset U$
are derived from the variation of degrees of $k_r$ freedom for every link $r\subset \tau$; the condition is 
$\hat{\theta}_{s(r, \nu), i}- \hat{\theta}_{s(r, \nu'), i}=0$ where the we have given $s(r, \nu)$ and $s(r, \nu')$ compatible orientations. 
Notice that the gluing equations correctly paste all wedges $s(\sigma, \nu)$ that meet at a co-dimension two simplex $\sigma$ interior to $U$; 
thus, the gluing field equations require that
\[
\forall \sigma \subset (U - \partial U) \quad 
\hat{\theta}_{s(\sigma, \nu), i} \mbox{ be independent of } \nu . 
\]

The momentum map for this system is given by $\Theta_L$, which gives us 
a version of Noether's theorem. 

The explicit calculation of the multisymplectic form, 
$\Omega_L(\tilde{v}_\xi, \tilde{w}_\eta, \tilde{A}(\tau_\nu))=
-\tilde{v}_\xi[\Theta_L(\tilde{w}_\eta, \tilde{A}(\tau_\nu)] 
+\tilde{w}_\eta[\Theta_L(\tilde{v}_\xi, \tilde{A}(\tau_\nu)] 
+\Theta_L([\tilde{v}_\xi, \tilde{w}_\eta], \tilde{A}(\tau_\nu))$, 
involves the derivative of $\hat{\theta}_{s, i}$. We obtain 
$\tilde{v}_\xi \hat{\theta}_{s, i} (\tilde{A}(\nu)) = \frac{\beta}{N}
(\xi_{r1} - \xi_{r2} - \hat{\xi}_{l2} + \hat{\xi}_{l1})^j \vartheta_{s, ij}$, where 
$\vartheta_{s, ij}= -\Rp \Tr(f_i f_j k_{r1} h_{l1} h_{l2}^{-1} k_{r2}^{-1})$ 
and $\hat{\xi}_l= h_{l1}^{-1} k_{r1}^{-1} \xi_l k_{r1} h_{l1}$. The multisymplectic form is 
\[
\Omega_L(\tilde{v} , \tilde{w} , \tilde{A}(\tau_\nu))=
\frac{\beta}{N}\sum_{r \subset \tau} 
\left[
\xi_{r}^i (-\eta_{r2} - \hat{\eta}_{l2} + \hat{\eta}_{l1})^j 
- 
\eta_{r}^i (-\xi_{r2} - \hat{\xi}_{l2} + \hat{\xi}_{l1})^j 
\right]
\vartheta_{s, ij} . 
\] 
Now we could write explicitly the conservation law implied by the multisymplectic formula. 
Given any first variations $v, w$ of any solution $A$ we have 
$\sum_{\partial U} \tilde{A}^\ast(\tilde{w} \lrcorner \tilde{v} \lrcorner\Omega_L) =0$. 

The reader could work out the geometric structure of Veselov's model from the action written in Subsection \ref{1dExamples} to discover many similarities with the structure presented above. Veselov's model is a discrete time integrable model; for a detailed study see \cite{Veselov, Veselov-Moser}

\noindent
{\em The reduced model} \\
In the scalar field the gluing conditions required that the increment in the field as one moved from atom $\nu$ to its neighbor $\nu'$ was divided into two equal increments 
$\phi_\tau - \phi_\nu = \phi_{\nu'} - \phi_\tau$, and this equation has a unique solution for $\phi_\tau$. In the present example the gluing condition, $\hat{\theta}_{s(r, \nu), i}- \hat{\theta}_{s(r, \nu'), i}=0$, is quite similar, and if we investigate only small angles it has a unique solution for $k_r$; that solution also satisfies the condition 
$\hat{g}_{s(r, \nu)} = \hat{g}_{s(r, \nu')}$, where 
$s(r, \nu)$ and $s(r, \nu')$ share the link $r$ and have compatible orientations. To be specific we define 
$\hat{g}_{s(r, \nu)} = k_{r} h_{l1} h_{l2}^{-1} k_{r2}^{-1}$ and 
$\hat{g}_{s(r, \nu')} = k_{r3} h_{l4} h_{l3}^{-1} k_{r}^{-1}$; then the solution is 
$k_r = \hat{g}_{s(r, \nu')} \hat{g}_{s(r, \nu)} (h_{l1} h_{l2}^{-1} k_{r2}^{-1})^{-1}$. 
The same condition can be used to eliminate all the $k$ variables of links that flow into a given $2$-cell $\sigma$. We can join all the wedges $\{ s_1, s_2, s_3 , s_4 \}$ 
which touch $C\sigma$ to form a cell dual to $\sigma$ and define 
$\hat{g}_{\sigma^\ast} = \hat{g}_{s_4} \hat{g}_{s_3} \hat{g}_{s_2} \hat{g}_{s_1}$ which is independent of all the $k$ variables of links that flow into $C\sigma$. 
After all those $k$ variables have been eliminated by solving the corresponding gluing  conditions, we obtain 
$\hat{g}_{s_i} = \hat{g}_{\sigma^\ast}^{\frac{1}{4}}$, and the action that describes this reduced model is 
\[
S_{\rm Euc}^r(A)= \beta
\sum_{\Box \subset U^r} 
4 [1 - \frac{1}{N} \Rp \Tr({g_{\Box}}^{\frac{1}{4}})] , 
\]
where the plaquettes, represented by the symbol $\Box$, are the minimal circuits of the lattice formed by the links that do not touch any $2$-cell and the region 
$U^r \subset U$ where the reduced action describes the model is the proper subset of $U$ composed of the union of cells dual to the vertices ($0$-cells) of the original cubical cellular decomposition that are interior to $U$. The relation between this action and Wilson's action is self-evident.

\noindent
{\em Can we build a reduced model of boundary data?}\\ 
The interior field equations $\sum_{s \supset l} \theta_{s, i}=0$ are not as simple to solve explicitly as the gluing equations were. 
In this case, then, proposing a new model that uses only $k_r$ variables corresponding to links in the boundary of atoms is more appropriate. 
There are several possibilities, and it would be attractive to study them given that they would be a classical counterpart of spin foam models for gauge fields defined in terms of the cell amplitudes \cite{Reisenberger94, ArocaGambiniFort}.

\subsection{Reisenberger's simplicial model for general relativity}
\label{MichaelsModel}
We apply the formalism of 
Subsections \ref{gauge} and \ref{BF+subsection} 
to a regularization of the Plebanski action for general relativity \cite{Plebanski} on a triangulated domain $U$ of spacetime. The action and the field equations of this model were written by Reisenberger; a detailed presentation of the model is given in 
\cite{Reisenberger}. Here we introduce the structural forms induced by the action. In particular, we give a multisymplectic formula for this model. In addition, the relation with the canonical framework developed in Section \ref{Canonical} shows how this covariant model is related to discrete hamiltonian frameworks based on data given in co-dimension one surfaces. Thus, this study contributes to the understanding of the relation between classical and quantum models of gravity at the discrete level. For a discussion of different aspects of the continuum limit see Section \ref{Reg-Coar-Cont} and \cite{Reisenberger}. 
We remark that this model belongs to a family of models that regularize modified BF theories, and the structures that we develop apply to all the models of this family. 

Reisenberger's model is the constrained $SU(2)$-BF theory 
based on the following action 
\[
S(\ae)= 
\sum_{\nu \subset U}[ \sum_{s\subset\nu} e_{s \, i} \theta_s^i - 
\frac{1}{60} \varphi^{ij}_\nu
\sum_{s, {s'}\subset\nu} e_{s \, i} e_{s' \, j} \sgn(s, {s'})] , 
\]
where $\theta_s^i = -2i\Tr(J^i g_{\partial s})$. 
We recall that the definition $g_{\partial s}= h_{l2}^{-1} k_{r2}^{-1} k_{r1} h_{l1}$ requires a choice of orientation for the wedge $s$, but since both $\theta_{s\, i}$ and $e_s^i$ change sign if the orientation of $s$ is flipped, the action is well-defined as it is written. 
We follow conventions 
where the Lie algebra elements are written as $\xi= \xi^k(i J_k)$ 
with the generators written in terms of Pauli's sigma matrices as 
$iJ_k = \frac{i}{2} \sigma_k$. 

Now we study the variation of action. Recall from Subsection \ref{BF+subsection} that in the local 1st order format a history is written as 
$\tilde{\ae}(\nu)=(\nu, \{ h_l \}_\nu , \{ k_r \}_\nu , \\ \{ e_s \}_\nu , \varphi_\nu)$ where $\varphi_\nu$ is a symmetric traceless matrix. 
For convenience variations are written in the form 
$\delta \tilde{\ae}(\nu)=\tilde{v}_\xi(\nu)=(\{ h_l \xi_l \in T_{h_l}SU(2) \}_\nu , 
\{ \xi_r k_r\in T_{k_r}SU(2) \}_\nu , 
\{ \xi_s \in T_{id}SU(2) \}_\nu , v_\nu \in V)$. 
While calculating the variation of the action we need the following derivatives:\\
(i)  
$e_{s \, i}  \frac{\partial}{\partial h_{l1}}
\theta_s^i dh_{l1}[h_{l1} \xi_{l1}] = e_{s \, i} 2\xi_{l1}^j \Tr(J_j J^i g_{\partial s})
= \xi_{l1}^j w_{s \, j}$, and\\
(ii) 
$e_{s \, i}  \frac{\partial}{\partial k_{r1}} \theta_s^i dk_{r1} 
[ \xi_{r1} k_{r1}]= e_{s \, i} 2 \xi_{r1}^j 
\Tr(J^i h_{l2}^{-1} k_{r2}^{-1} J_j  k_{r1} h_{l1}) 
= \hat{\xi}_{r1}^j w_{s \, j}=  \xi_{r1}^j u_{\sigma, j}$,\\ 
where we have defined 
$\xi_{r1} = k_{r1} h_{l1} \hat{\xi} h_{l1}^{-1} k_{r1}^{-1}$ and $u_{\sigma \, j}= k_{r1} h_{l1} w_{s \, j} h_{l1}^{-1} k_{r1}^{-1}$. 
A short calculation shows that if the orientation of $\sigma$ is reversed (which implies that the orientation of $s$ is reversed and that the parallel transports from $C\nu$ to $C\sigma$ used to define $w_s$ follow the other path), then the new variable changes sign 
$u_{\bar{\sigma}}= -u_{\sigma}$. 
Now we write the derivative of the action as 
$dS(\ae) [v] =
- \sum_{U -\partial U}
\tilde{\ae}^\ast(\tilde{v} \lrcorner \hat{\Omega}_L) 
+ 
\sum_{\partial U} 
\tilde{\ae}^\ast(\tilde{v} \lrcorner \Theta_L)$, with 
\[
\Theta_L(\tilde{v}_\xi , \tilde{\ae}(\tau_\nu)) = 
\tilde{v}_\xi L(\tilde{\ae}(\nu))|_\tau =
\sum_{r \subset \tau} \xi_{r}^j u_{\sigma \, j} , 
\]
\begin{eqnarray*}
\hat{\Omega}_L(\tilde{v} , \tilde{A}(\nu)) &=&
-\tilde{v}_\xi L(\tilde{A}(\nu)) = 
-\sum_{l \subset \nu} \xi_{l}^j \sum_{s \supset l} w_{s\, j} 
-\sum_{r \subset \nu} \xi_{r}^j u_{\sigma \, j} \\
&-&\sum_{s\subset\nu} \xi_{s \, i} (\theta_s^ i -\frac{1}{60} \varphi^{ij}_\nu 
\sum_{{s'}\subset\nu} e_{s' \, j} \sgn(s, {s'}))\\
&+&\frac{1}{60} v^{ij}_\nu
\sum_{s, {s'}\subset\nu} e_{s \, i} e_{s' \, j} \sgn(s, {s'})] . 
\end{eqnarray*}
The field equations follow from the bulk terms in the variation of the action, $-\sum_{U -\partial U} \tilde{\ae}^\ast(\tilde{v} \lrcorner \hat{\Omega}_L)$. 
These equations were written by Reisenberger, and a detailed exposition is given in \cite{Reisenberger}. Here we only give brief comments. 
If our region of interest is a single atom the extremum condition would not involve any derivatives with respect to the $k_r$ degrees of freedom because they are ``boundary degrees of freedom''; the rest of the terms in $\hat{\Omega}_L(\tilde{v} , \tilde{A}(\nu))$ are required to vanish for any variation for $\ae$ to be a solution. 
If our region contains several atoms glued along shared faces, the gluing field equations consist purely of derivatives with respect to $k_r$ degrees of freedom for $r\subset \tau$. 
Note 
that the $k_r$-gluing 
equation is a condition on an object, $u_{\sigma \, j}$, 
associated with $\sigma$ (the co-dimension two simplex where $r$ finishes) that must be shared by the two atoms containing $r$. Since this condition must be met by every pair of neighboring atoms that intersect $\sigma$, 
the object $u_{\sigma \, j}$ is independent of the atom used to calculate it. 

% New
The boundary degrees of freedom at $\partial\nu$ are associated to the graph formed by its boundary links $\Gamma_{\partial \nu}$. There is phase space associated to that graph 
$\times_{r \subset \Gamma_{\partial \nu}} T^\ast SU(2) \ni 
\{(k_r, u_{\sigma(r)}) \}_{r \subset \Gamma_{\partial \nu}}$. 
There is more economical description where the associated graph 
$\tilde{\Gamma}_{\partial \nu}$
does not have bivalent vertices. The configuration variables are $M_{ij}= k_{rj}^{-1} k_{ri}$ describing the parallel transport from $C\tau_i$ to $C\tau_j$. The corresponding momentum variable is $u_\sigma$ with the appropriate orientation and parallel transported from $C\sigma$ to either $C\tau_i$ (giving $E_{ij}$) or to $C\tau_j$ (giving $E_{ji}$). The momentum variables defined in this way satisfy the relation $E_{ij} = - M_{ij} E_{ji} M_{ij}^{-1}$. 
%We remark that not only the variables match, but also the symplectic potential is 
%correctly implemented because the definition of momenta already contains the 
%appropriate information of the lagrangian. 
Points in the phase space associated to $\tilde{\Gamma}_{\partial \nu}$ can be associated a twisted geometry \cite{twisted} or a spinning geometry \cite{Spinning}; one of the field equations given above implies that the sum of the $E$s associated to a codimension $1$ face vanishes, and this becomes an essential ingredient of the geometrical interpretation. 
In the spacetime spirit advocated in this article it would be natural to assign a continuous geometry to the the boundary of a spacetime atom $\partial \nu$. It would also be natural to look for $4$-geometries associated to histories (or alt least to solutions) at $\nu$ which are compatible with any $3$-geometry associated to $\partial \nu$. Thus, twisted and spinning geometries are relevant to Reisenberger's model, but there may be better suited models for the geometry of spacetime atoms and their boundary.
%An intrinsic difficulty for finding a geometrical model for the atom is that the 
%connection should not be flat inside an atom, leaving only the possibility of 
%postulating a notion of a connection with ``constant curvature'' which is only obvious 
%in the case of abelian groups. 
%

Comparing this model with our discrete time model for rigid body motion presented in Subsection \ref{1dExamples} is illuminating. In that model the variables corresponding to the body angular momentum play an analogous role to the variables $u, w$ of Reisenberger's model. In fact, we have adapted the notation in our model for rigid body motion to facilitate the comparison.  

The structural form $\Theta_L$, providing a momentum map and a version of Noether's theorem, 
is exceptionally simple; all the complications inherent to general relativity are deposited in finding solutions to the ``bulk field equations,'' while gluing is kept simple --as simple as in a model for BF theory. 
The explicit expression for the multisymplectic form 
$\Omega_L(\tilde{v}_\xi , \tilde{w}_\eta , \tilde{\ae}(\tau_\nu)) =
-\tilde{v}_\xi \Theta_L(\tilde{w}_\eta , \tilde{\ae}(\tau_\nu)) 
+\tilde{w}_\eta \Theta_L(\tilde{v}_\eta , \tilde{\ae}(\tau_\nu)) 
+ \Theta_L([\tilde{v}_\xi , \tilde{w}_\eta ] , \tilde{\ae}(\tau_\nu))$ is 
\[
\Omega_L(\tilde{v}_\xi , \tilde{w}_\eta , \tilde{\ae}(\tau_\nu)) = 
-\sum_{r \subset \tau} ( \eta_{r}^j \tilde{v}_\xi(u_{\sigma \, j}) 
- \xi_{r}^j \tilde{w}_\eta(u_{\sigma \, j})) (\tilde{\ae}(\nu)) . 
\]
Then, the multisymplectic formula 
$\sum_{\partial U} \tilde{\ae}^\ast(\tilde{w} \lrcorner \tilde{v} \lrcorner\Omega_L) =0$ holds for any solution $\ae$ and any two first variations $v, w$ of it.

In the previous examples we have commented on the possibility of solving the gluing equations to derive a reduced model. In this model, we can solve the gluing equations as conditions on the variables $e_s$, 
obtaining a system whose histories are described by the same variables as Reisenberger's model except that it uses 
one single variable $u_\sigma$ per co-dimension two simplex $\sigma$ instead of one variable per wedge $s\subset \sigma^\ast$. For most reduced histories one would be able to undo the reduction and find one history of Reisenberger's model corresponding to he original reduced history. 
One could try to reduce it further and eliminate the $\{ k_r \}$ variables as we did in the lattice gauge theory model; however, there is no simple expression of the reduced action in which the $\{ k_r \}$ variables drop out. 

An alternative reduction would start with the original model and solve the interior field equations of an atom $\nu$ 
to eliminate as much as possible of 
the $\{ h_l \}$ variables, the $\varphi$ variable and the $\{ e_s \}$ variables. 
A further alternative defines a simple model in terms of boundary variables directly. 
These two options for classical models 
have not been sufficiently explored, and their quantum analogs write the amplitude of a history as a product of amplitudes associated with the connection on the boundaries of spacetime atoms, as originally proposed by Reisenberger \cite{Reisenberger94}. 
%A very interesting recent proposal by Neiman is using null 
%normals to codimension two faces to parametrize solutions, 
%and write an action directly in terms of such unconstrained variables \cite{Neiman}. 
%

\section{Regularization, coarse graining and\\
continuum limit}
\label{Reg-Coar-Cont}

\noindent
{\em Decimation}\\ 
% - Coarse graining by decimation \pi_\Delta 
Our framework rests on a decimation 
map $\pi_\Delta$ that produces sections in the discrete framework from sections in the continuum framework. 

\noindent
{\em Regulariation}\\ 
% - Reg brings questions to scale Delta (i_\Delta^\ast), i section of \pi 
Regularization plays two roles. The first one is to 
bring questions to scale $\Delta$ from the continuum; since the regularization map acts on functions, here we will write it as a pull back map $i_\Delta^\ast$, where 
$i_\Delta^\ast \circ \pi_\Delta^\ast = id$. 
The second job of regularization is to give us a discrete action from the action in the continuum. 
% - We do not commit to a specific regularization scheme 
In this work we do not commit to a particular regularization scheme. 

% - Hodge dual and laplacian OK in structured atoms
The structured discretization used in this article allows for a definition of Hodge dual for cochains and a related definition of a laplacian along the lines described in \cite{LatticeRegularization}. In the language used in those references, 
the structured regularization used in our work includes the essential ingredients of the prime cellular decomposition and of the dual cellular decomposition unifying them into a single entity with (structured) atoms. Thus, the important body of work contained in  \cite{LatticeRegularization} can be used, with small adjustments, to produce models within our framework. 

\noindent
{\em Continuum limit of field equations and geometric structures}\\ 
We developed geometric structures related to a discretization that traded 
partial differential equations for difference operators. 
A natural question is:\\ 
Does a continuum limit reproduce partial differential equations from 
our difference operators, and does it yield limiting structures matching the corresponding structures in the continuum? \\
Veselov studies this issue and concludes that, the structure of some of his models 
converges to the continuum structures \cite{Veselov}. 
For our discrete field theories we can consider the same continuum limit, and study its convergence: 
In the case of the scalar field, we fix a smooth section $\phi$ and a variation of it $v = \delta \phi$; then we decimate them to each of a sequence of refining scales $\{ \Delta_n \} \to M$ 
producing $\{ \phi_n \}$, $\{ v_n \}$. The limit to be studied is 
\[
\lim_{\Delta_n \to M} d S_{\Delta_n} (\phi_n) [ v_n] . 
\]
Convergence to the appropriate continuum limit implies the convergence of both 
\[
\sum_{U - \partial U} \tilde{\phi}_n^\ast ( \tilde{v}_n \lrcorner \hat{\Omega}_{L_n})
\quad \to \quad 
\int_U j^1{\phi}^\ast ( j^1{v} \lrcorner \hat{\Omega}_{\cal L}) 
\, \mbox{ and}
\]
\[
\sum_{\partial U} \tilde{\phi}_n^\ast ( \tilde{v}_n \lrcorner \Theta_{L_n})
\quad \to \quad 
\int_{\partial U} j^1{\phi}^\ast ( j^1{v} \lrcorner \Theta_{\cal L}) . 
\]
The continuum limit of all the models that we presented reproduces the original continuum model in this sense. 
The models that require a more delicate analysis are the lattice gauge theory model and Reisenberger's simplicial gravity model; the continuum limit of our lattice gauge theory model can be studied following the route usually followed for 
Wilson's lattice gauge theory \cite{Creutz}, and the continuum limit of Reisenberger's model is carefully studied 
in \cite{Reisenberger}. 

\noindent
{\em Continuum limit of solutions}\\ 
A more ambitious objective is the study of convergence of solutions in a continuum limit. 
The first step is to fix boundary conditions (or initial conditions) in the continuum 
$\psi^c$ 
and coarse grain them to 
a refining sequence of scales obtaining the corresponding conditions at each scale 
$\{ \psi_n \}$. The second step is to 
find a solution at each scale $\{ \phi_n(\psi^c) \}$ and bring it to the continuum using the map $i_n$. Clearly, finding solutions is a very delicate issue in which the details of the discrete model have an impact. (For studies related to this issue see 
\cite{GambiniETAL} and \cite{DittrichETAL}.) The third step is to study the limit 
$\lim_{\Delta_n \to M} i_n \phi_n(\psi^c)$. 
Alternatively, we can study the continuum limit using evaluations of a family of observables from the continuum 
\[
\lim_{\Delta_n \to M} i_n^\ast f^c(\psi_n) . 
\]
Quantum lattice field theories give predictions about observables from the continuum following the same strategy; 
at each scale a measure is constructed (after adjusting the coupling constants) and 
the convergence of the $n$-point functions (or the Wilson loops) is studied \cite{Creutz}. 
In this respect, loop quantum field theories have the interpretation of being the continuum limit constructed from effective theories in exactly this way \cite{LQasContLim}.

\noindent
{\em Correction of the structures at a given scale}\\ 
Two different scales may be related by coarse graining (by decimation in our case).  
Since our decimation preserves locality, the map 
can be written in the local first order format. Consider an atom of scale $\Delta$ written 
in scale $\Delta'$ as a chain of atoms $\nu= \nu'_1+ ... + \nu'_m$. In the case of the scalar field we require that there is a 
$\Delta'$-atom $\nu'_\star$ such that  $C\nu= C\nu'_\star$, and that in the decomposition of $\partial \nu$ at scale 
$\Delta'$ we have boundary faces $\{ \tau'_\star \}_{\tau \subset \partial \nu}$ such that 
$\{ C\tau = C\tau'_\star \}_{\tau \subset \partial \nu}$. In this situation, 
the coarse grained history $\tilde{\phi}(\nu)=(\pi_{\Delta \Delta'} \tilde{\phi'})(\nu)$ 
is defined as 
\[
(\pi_{\Delta \Delta'} \tilde{\phi'})(\nu) = (\nu, \phi_\nu= \phi'_{\nu'_\star} , \{ \phi_\tau= \phi'_{\tau'_\star} \}) . 
\]
In the case of gauge degrees of freedom, where measuring at scale $\Delta$ means selecting a collection of paths in spacetime, the condition necessary to have a coarse graining map by decimation is that each path in the mentioned collection of paths at scale $\Delta$ 
is a composition of 
paths of scale $\Delta'$; the rest of the paths in $\Delta'$ are ignored. Then we define 
\[
(\pi_{\Delta \Delta'} \tilde{A'})(\nu) = (\nu, \{ h_l = h_{l' \, n_l} \circ ... \circ h_{l' \, 1} \}_\nu , 
\{ k_r = k_{r' \, n_r} \circ ... \circ k_{r' \, 1}\}_\nu) . 
\]
The relation written above refers to a case in which the $l$ links of scale $\Delta$ are compositions of $l$ links of scale $\Delta'$, and  
$r$ links of scale $\Delta$ are compositions of $r$ links of scale $\Delta'$. It is always possible to refine a cellular decomposition in such a way that this property holds. The formalism can also deal with cases in which the $l$ links become compositions of $l$ and $r$ links. 

When a solution in a fine scale 
is coarse grained it goes to a history which in general is not a solution; the 
multisymplectic formula and Noether's conservation law do not hold on it. 
The action at the coarse scale can be corrected \cite{Perfect}, 
which corrects the equations of motion 
and the geometric structures fixing the incompatibility mentioned above. 
This is the classical analog of coarse graining the measure in statistical field theory and 
quantum field theory. The procedure for correcting the action is  the following: 
Given a state $\phi$ at scale $\Delta$, 
the action at scale $\Delta'$ poses a variational problem among the states at scale 
$\Delta'$ that satisfy $\pi_{\Delta \Delta'} \phi' = \phi$. If $\phi'$ is a minimum of the mentioned variational problem, the value of the corrected action at scale $\Delta$ on $\phi$ is defined to be 
\[
S_{U,\Delta (\Delta')} (\phi) = S_{U,\Delta'}(\phi') . 
\]
There are several examples where this strategy has been successful. An interesting explicit example is the implementation of this program for a reparametrization invariant path integral \cite{Bahr:2011uj}.

In 1-dimensional spacetimes, 
the boundary data of each atom $\nu$ at scale $\Delta$ 
provides boundary data at the finer scale $\Delta'$ that may determine a solution. If we consider the variational problem with this boundary data at finer and finer scales, $S_{\nu, \Delta (\Delta_n)}(\phi)$ may converge to the continuum value 
$S_\nu (\phi)$, which is Hamilton's principal function for the given boundary data. 

On the other hand, if we follow the same procedure in spacetimes of higher dimensions, 
the data on $\partial \nu$ available at scale $\Delta$ is not enough to determine 
solutions inside $\nu$ for models defined at finer scales. 
Boundary data $\psi$ at scale $\Delta$ 
determines an ensemble of $\Delta'$ boundary data, $\pi_{\Delta \Delta'}^{-1}(\psi)$. 

The obvious draw back of working with corrected actions to achieve compatibility of different scales is the need of solving, 
or partially solving, the variational problem. 
This should not arise as a surprise, since the incompatibility that motivated correction involves two scales related by coarse graining and coarse grained {\em solutions} at the fine scale. 
On the other hand, we had found compatibility at the level of the field equations and 
the geometric structures (in the continuum limit). 

In the quantum theory one can use coarse graining map to define a continuum limit through 
an inverse limit. This continuum limit is related to the continuum limit of solutions described above; a description of both limits and their relation in the terminology used in this work appears in \cite{LQasContLim}.

\section{Outlook}
% 8) Outlook: what can be done in the short term with this formalism
% - Continuum / macroscopic limit and semiclassical limit
% - Quantization will strengthen the physical interpretation of SF models
\label{summary}
Our formalism provides a classical counterpart to spin foam models, in which spacetime 
atoms $\nu$ are autonomous minimal regions of spacetime that enjoy a complete version of the formalism. 
In our framework, the variation of the action decomposes into bulk and boundary terms 
$dS(A) [v] =
- \sum_{U -\partial U}
\tilde{A}^\ast(\tilde{v} \lrcorner \hat{\Omega}_L) 
+ 
\sum_{\partial U} 
\tilde{A}^\ast(\tilde{v} \lrcorner \Theta_L)$ 
giving origin to geometrical structure. First of all, the form $\Theta_L$ provides a covariant momentum map linking symmetries and conserved quantities, and it generates the multisymplectic form $\Omega_L = -d\Theta_L$ which is conserved in covariant evolution. 
On the other hand, the form $\hat{\Omega}_L$ is responsible for the equations of motion in the interior of spacetime atoms and for the simple gluing conditions that let us find solutions on larger regions of spacetime amalgamating solutions over atoms. 
The clean separation between bulk and boundary parts makes gluing transparent: neighboring regions share boundary data and gluing conditions asking for momentum matching appear when regions are fused. 
These structures have (some times implicit) quantum counterparts in spin foam models. 
We hope that our contribution, making available 
the classical counterparts of such structural properties, makes it possible for some research lines to advance further or enter a new level. One example is the study of the semiclassical limit of quantum gravity spin foam models, in which most of the research has been carried on at the level of a single atom; this study may now progress to study larger regions of spacetime aided with compatible the structures of quantum and classical gluing. 

In the case of lattice gauge theories, the transfer matrix for lattice gauge theory can be constructed gluing cellular propagators (or ``atomic propagators'' in the terminology used here); in this article we wrote the discrete field equations for the classical atomic propagator which is the classical counterpart of the mentioned atomic propagator. 
We presented a lagrangian picture and a hamiltonian picture 
with their respective structures and compatibility with the mentioned classical cellular propagator. 
Along this line, 
comparing our formalism with the classical hamiltonian formalism of Kogut and Susskind \cite{K-S}
would be interesting. 
Another line suggested by our framework is 
investigating 
the classical counterpart of the heat kernel action, and study the resulting theory because it is the most elegant spin foam model for lattice gauge theory \cite{Reisenberger94}. 

With quantum gravity in mind, the first thing to mention is that most of our study of Reisenberger's model in Subsection \ref{MichaelsModel} would carry over to any other model with similar degrees of freedom and gauge symmetries. 
The model's action determines a covariant Legendre transform bringing the canonical structure of the discrete dual jet bundle to the lagrangian setting. 
Since the boundary variables in each atom are $SU(2)$ group elements assigned to every boundary link, in the canonical side 
the atomic phase space boundary data consist of a collection of cotangent bundles $T^\ast SU(2)$; 
in the discussion of Reisenberger's model we showed how to relate this data to data for twisted geometries \cite{twisted} and spinning geometries \cite{Spinning}. 
For solutions to the field equations we have the conservation law given by the multisymplectic formula linking the pieces of $3$-geometry at the boundary of a region. 
The clear question is what is the 4-geometry corresponding to histories (or at least to solutions) compatible with a given $3$-geometry at the boundary?

Does Regge calculus (written in appropriate variables) fit as one of the examples of 
the ``atomic boundary data'' formalism presented in this article? 
We think that this project may not be a difficult one, given that Regge calculus is a 
well-studied subject with many previous and recent results. We should mention that the work of Dittrich and Höhn \cite{D-H} 
studies Regge calculus from a covariant hamiltonian view point. Their procedure uses a slightly different language, but their results could be described as including 
an implementation of what we describe in Section \ref{Canonical} mixed with Sorkin's method for constructing solutions 
\cite{Sorkin:1975ah}. 

Since our framework includes tools to study coarse graining, this opens another avenue of study. The study of classical coarse graining may also help us get a sharper understanding of coarse graining of quantum models and the study of a macroscopic limit.  

Let us also comment on the possibility of using this finite-dimensional discrete framework to investigate a covariant version of geometric quantization. Here we have a family of classical models and a corresponding family of quantum models waiting for the quantization problem to be clearly stated and tackled. In this respect, the presence of the General Boundary Field Theory formalism \cite{GBFT} 
provides the conceptual framework needed for this enterprise. We consider that the development of geometric quantization along these lines would provide a more complete understanding of quantum physics within the GBFT perspective. 

Finally, we comment on the comparison between our framework and previous approaches to discrete multisymplectic field theory. With the objective of comparing those frameworks, we developed the reduced formalism presented in Section \ref{Reduced}. We showed that the formalisms are equivalent in the sense of the relation between their spaces of solutions. 
The main disadvantage of reduced models as compared to the original models on a structured discretization 
is that the geometric structure is not simple any more. Recall that in our original framework the geometric structure was independent of the potential. The conservation laws had the same form as those of a free theory, while complications were secluded to solving the field equations. 
In contrast, 
for the reduced system 
the expressions for the conservation of the symplectic structure and for Noether's theorem may be complicated. Given that the main importance of conservation laws is to give information about the system even when solving the field equations is difficult or impossible, having simple expressions for the conservation laws is a significative advantage of our non-reduced framework. 

We hope that the simplicity of our conservation laws is used in future numerical applications.

\section*{Acknowledgements} 
This work was partially supported by grant Conacyt-80118. We thank the referees for their comments and suggestions.  

%\section*{References}

\end{document}